\documentclass[
  aps,pre,reprint,superscriptaddress,
  amsmath,amssymb,nofootinbib,nobibnotes
]{revtex4-2}

\usepackage[T1]{fontenc}
\usepackage[utf8]{inputenc}
\usepackage[english]{babel}
\babelprovide[import=en-US]{English (US)}
\usepackage{graphicx,bm,physics}
\usepackage{booktabs,supertabular,placeins,comment,xcolor,etoolbox}

\newcommand{\M}{\mathsf{M}}
\newcommand{\A}{\mathsf{A}}
\newcommand{\B}{\mathsf{B}}
\newcommand{\X}{\mathsf{X}}
\newcommand{\II}{\mathsf{I}}
\newcommand{\xbar}{\bar{x}}

\makeatletter
\let\APS@bibliography\thebibliography
\let\APS@endbibliography\endthebibliography
\makeatother
\usepackage[resetlabels]{multibib}

\makeatletter
\def\@extra@binfo{SI}
\newcites{SI}{SI References}
\def\@extra@binfo{}

\AtBeginDocument{%
  \edef\pre@bibdata{\jobname\bibdata@app}%
  \let\thebibliography\APS@bibliography
  \let\endthebibliography\APS@endbibliography
  \def\NAT@bibsetnum#1{%
    \setlength{\topsep}{0pt}\NATx@bibsetnum{#1}}%
}

\newenvironment{SIbibliography}{%
  \def\@extra@b@citeb{SI}%
  \let\SI@original@citex\@citex
  \def\@citex{\let\@auxout\@auxoutSI\SI@original@citex}%
  \let\SI@original@nocite\nocite
  \renewcommand{\nocite}[1]{{%
    \let\@auxout\@auxoutSI\SI@original@nocite{##1}}}%
  \patchcmd{\endthebibliography}
    {\label{LastBibItem}}{\label{LastSIBibItem}}{}{}%
  \immediate\write\@auxoutSI{\string\citation{apsrev42Control}}%
}{}
\makeatother

\usepackage[colorlinks=true,linkcolor=blue,citecolor=blue,urlcolor=blue]{hyperref}

\renewcommand{\thesection}{\Roman{section}}

\begin{document}

\title{Linear response in generalized Lotka--Volterra systems with a finite number of species}

\author{Koya Katayama}
\email{koya.katayama@ubi.s.u-tokyo.ac.jp}
\affiliation{Department of Physics, The University of Tokyo, 7-3-1 Hongo, Bunkyo-ku, Tokyo 113-0033, Japan}

\author{Sosuke Ito}
\affiliation{Department of Physics, The University of Tokyo, 7-3-1 Hongo, Bunkyo-ku, Tokyo 113-0033, Japan}
\affiliation{Universal Biology Institute, The University of Tokyo, 7-3-1 Hongo, Bunkyo-ku, Tokyo 113-0033, Japan}

\begin{abstract}
Understanding how species interactions shape ecosystem responses to perturbations is a central challenge in ecology. Random matrix theory (RMT) has been widely used to predict the responses of large random ecosystems, but its predictions may not accurately describe individual finite ecosystems. Here, we derive rigorous geometric bounds on the numerical ranges and spectra of Jacobian matrices that are valid for any finite number of species without assuming randomness. 
These bounds link long-term stability and short-term stability to the symmetric component of abundance-weighted interactions, and constrain the maximum oscillation frequency by the skew-symmetric component.
Applied to random ensembles, the bounds also yield constraints on ensemble averages of long-term stability, short-term stability, and the maximum oscillation frequency in terms of standard RMT parameters. 
Using empirical data from marine food webs, we show that our bounds provide informative constraints on short-term stability and the maximum oscillation frequency even where RMT-based predictions fail. These results establish structural constraints on local ecosystem responses that complement RMT-based statistical predictions for large random ecosystems.
\end{abstract}
\maketitle

\section{Introduction}
Understanding the response of ecosystems to environmental perturbations has long been a central question in ecology~\cite{pimm1984complexity, takeuchi1996global, may2001stability, landi2018complexity, CHEN20241, arnoldi2016resilience, allesina2015stability}. In particular, random matrix theory (RMT) has provided a widely used framework for studying the linear response of ecological communities described by the generalized Lotka–Volterra (gLV) equation~\cite{allesina2015stability}. In this framework, Jacobian matrices evaluated at fixed points are modeled as random matrices, and their spectral properties are related to statistical features of species interactions. Following Gardner and Ashby~\cite{gardner1970connectance}, May~\cite{may1972will} established a trade-off between linear stability and complexity in random ecological communities using RMT~\cite{mehta1967random, ginibre1965statistical, Girko1984, bai1997circular, edelman1997probability, tao2010random}. Subsequent work extended this framework to incorporate ecological structure and different interaction types~\cite{allesina2012stability, tang2014correlation, tang2014reactivity, mambuca2022dynamical, yang2023reactivity, valigi2024local}, building on further developments in RMT~\cite{Girko1985, sommers1988spectrum, naumov2012elliptic, Nguyen2015, orourke2014low}.

However, applying RMT-based predictions to individual ecosystems requires caution. Widely used RMT predictions describe asymptotic behavior in the limit of infinitely many species and need not accurately characterize finite ecological networks. Moreover, the assumptions about randomness and correlations among matrix elements may not capture the interaction structure of real ecosystems. Empirical studies have reported weak or absent relationships between complexity and stability~\cite{james2015constructing, jacquet2016no}, highlighting limitations of directly applying standard RMT predictions to real systems. These considerations motivate complementary approaches that provide rigorous constraints on the linear response of finite ecological communities without assuming an infinite number of species.

Here, inspired by recent work on spectral and numerical-range constraints in stochastic thermodynamics~\cite{barato2017coherence, uhl2019affinity, oberreiter2022universal, ohga2023thermodynamic, xu2025thermodynamic, kolchinsky2026cycle}, we derive geometric bounds on the numerical range and spectrum of the Jacobian matrix. These bounds hold for any finite number of species without assuming randomness. 
In particular, these geometric constraints provide upper bounds on three quantities: the spectral abscissa, which determines long-term stability; the numerical abscissa, or reactivity, which characterizes short-term stability; and the maximum oscillation frequency of linear oscillatory modes. The bounds on the spectral and numerical abscissas depend on the symmetric component of abundance-weighted interspecific interactions together with intraspecific effects, whereas the bound on the maximum oscillation frequency depends on the skew-symmetric component.
Because the geometric bounds hold for arbitrary finite-dimensional real square matrices, they also yield alternative spectral bounds through similarity transformations associated with different scalings of perturbations.

Applying our results to random ensembles further yields bounds on ensemble averages of the spectral abscissa, reactivity, and the maximum oscillation frequency expressed in terms of the same parameters used in standard RMT predictions. These inequalities remain valid for every finite community size. Although they share some qualitative parameter dependence with RMT predictions, their scaling with species number differs, and the bounds can become increasingly loose as the number of species grows. Our approach therefore complements RMT by providing rigorous constraints on individual finite systems and their ensemble averages, while RMT characterizes the typical asymptotic behavior of large random communities.

We assess the empirical usefulness of our framework using Jacobian matrices reconstructed from 232 food-web models. Both our bounds and the RMT-based estimates capture overall trends across these systems, particularly for reactivity. However, the RMT predictions examined here tend to overestimate the spectral abscissa and underestimate reactivity, while estimates of the maximum oscillation frequency exhibit substantial scatter and outliers. Our bounds are satisfied for every system analyzed and provide informative constraints on reactivity and the maximum frequency of linear oscillatory modes. The bound on the spectral abscissa is less tight, reflecting the potentially large gap between the spectral and numerical abscissas. Nevertheless, the spectral abscissa and its upper bound are positively correlated across systems. These findings demonstrate how geometric constraints valid at finite community size can complement RMT predictions in characterizing the local responses of ecological networks.

\section{Background}
\subsection{Model}
To discuss the linear response of ecosystems, we consider the generalized Lotka--Volterra (gLV) equation, a standard model in theoretical ecology.
Consider an ecosystem consisting of $N$ species.
Let $x_i(t)\ge 0$ denote the abundance of species $i\in\{1,\ldots, N\}$ at time $t$.
The gLV equation describes the dynamics of $\bm{x}(t):=(x_1(t),\ldots,x_N(t))^\top$:
\begin{align}\label{eq:gLV}
    \frac{d}{dt}x_i(t)=\left(r_i+\sum_jA_{ij}x_j(t)\right)x_i(t),
\end{align}
which generalizes the original Lotka--Volterra equation~\cite{lotka1920analytical, volterra1926}.
Here, $r_i$ is the intrinsic growth rate of species $i$, and $A_{ij}$ is the interaction strength from species $j$ to species $i$.
The matrix $\A:=(A_{ij})$ is called \textit{the interaction matrix}.
Three representative interaction types are (1) $A_{ij}>0$ and $A_{ji}>0$ (cooperative), (2) $A_{ij}<0$ and $A_{ji}<0$ (competitive), and (3) $A_{ij}>0$ and $A_{ji}<0$ (or $A_{ij}<0$ and $A_{ji}>0$) (predator--prey) (see Fig.~\ref{fig:non_normal}A).
As shown below, $\A$ determines the Jacobian matrix at the fixed point.

Since our interest is the linear response near the fixed point, we linearize \eqref{eq:gLV}.
The fixed point of \eqref{eq:gLV}, denoted by $\bar{\bm{x}}$, satisfies $(r_i+\sum_jA_{ij}\xbar_j)\xbar_i=0$.
Assuming $\xbar_i>0$ for all $i$, the perturbation vector $\delta\bm{x}(t):=\bm{x}(t)-\bar{\bm{x}}$ obeys
\begin{align}\label{eq:delx}
    \frac{d}{dt}\delta\bm{x}(t)=\M\delta\bm{x}(t)
\end{align}
with $(\M)_{ij}:=\xbar_iA_{ij}$.
As discussed next, the linear response around the fixed point is determined by properties of $\M$.

\subsection{Characteristics of linear response}
In linear stability analysis, we can discuss the local stability around a fixed point based on the eigenvalues of $\M$.
Long-term stability is quantified by \textit{the spectral abscissa} $\alpha:=\max\{\Re\lambda\mid\lambda\in\Lambda(\M)\}$, where $\Lambda(\M)$ denotes the spectrum of $\M$.
Since $\|\delta\bm{x}(t)\|\sim e^{\alpha t}$ for large $t$, $\alpha<0$ implies that $\|\delta\bm{x}(t)\|$ eventually vanishes.
Thus, $\alpha$ determines long-term stability.

While long-term stability is quantified by the spectral abscissa, short-term stability is quantified by \textit{the numerical abscissa}, defined via \textit{the numerical range}.
The numerical range of $\M$ is $W(\M):=\left\{\bm{u}^\dagger\M\bm{u}\;\middle|\; \bm{u}\in\mathbb{C}^N,\, \bm{u}^\dagger\bm{u}=1\right\}$.
The numerical abscissa of $\M$ is $\omega:=\sup\{\Re \lambda\mid \lambda\in W(\M)\}$, also called \textit{the reactivity}~\cite{neubert1997alternatives, tang2014reactivity}.
Since $\left.\frac{d}{dt}\|\delta\bm{x}(t)\|\right|_{t=0}\le \omega\|\delta\bm{x}(0)\|$~\cite{trefethen2020spectra}, $\omega$ determines short-term stability.

\begin{figure}[htbp]
    \centering
    \includegraphics[width=\linewidth]{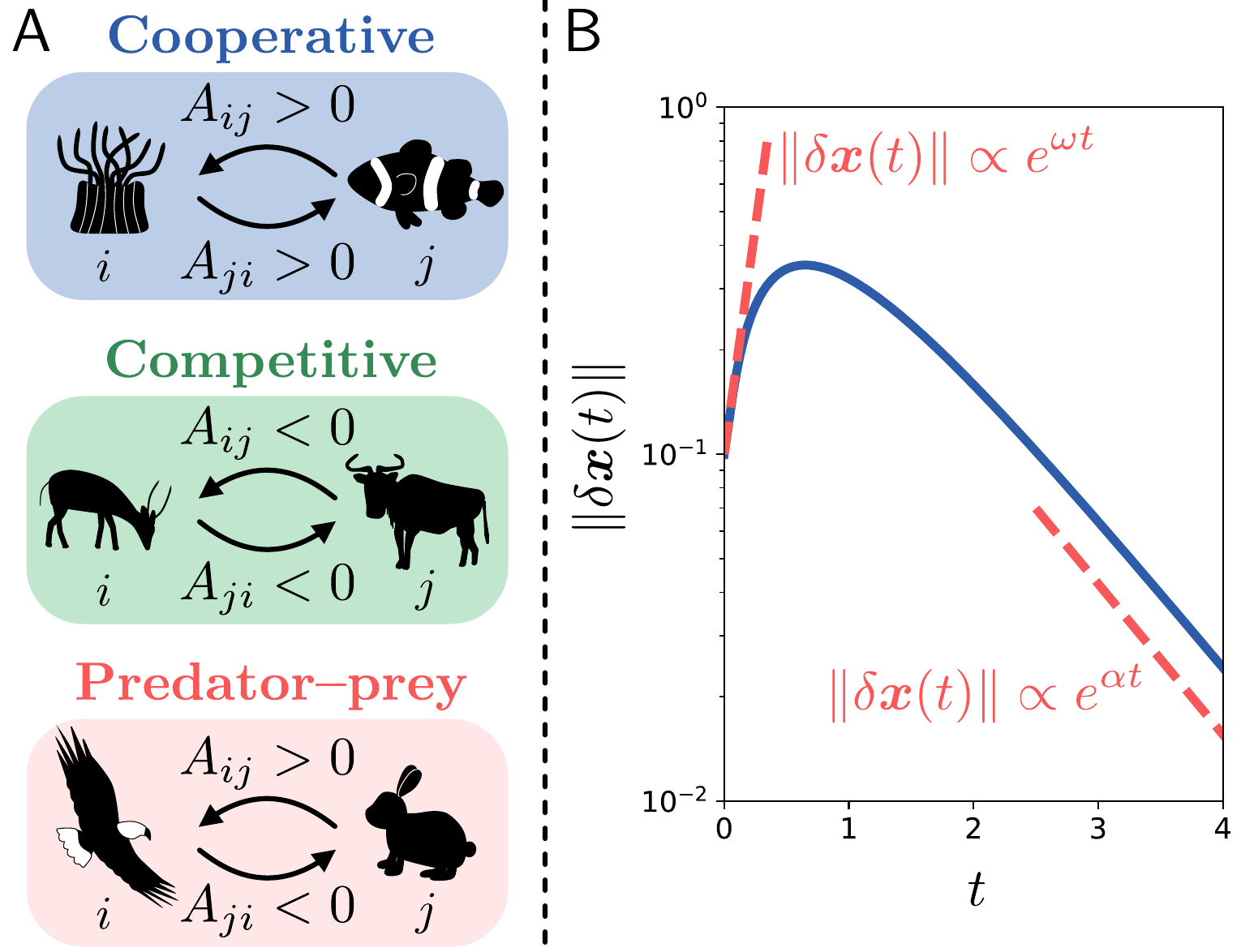}
    \caption{(A) Three types of interaction between species in terms of $\mathsf{A}=(A_{ij})$.
    (B) The behavior of $\|\delta\bm{x}(t)\|$ around the fixed point. The behavior is determined by both the spectral abscissa $\alpha$ and the numerical abscissa $\omega$. The solid line shows $\|\delta\bm{x}(t)\|$ on a logarithmic scale.
    The two dashed lines indicate proportionality to $e^{\omega t}$ and $e^{\alpha t}$, respectively.
    See SI Appendix for details of the numerical calculation to generate this plot.}
    \label{fig:non_normal}
\end{figure}

We now relate the spectral abscissa and the numerical abscissa.
In general, $\alpha\le\omega$ holds because $\Lambda(\M)\subseteq W(\M)$~\cite{trefethen2020spectra}.
The equality $\alpha=\omega$ holds if (but not only if) $\M$ is normal~\cite{trefethen2020spectra}.
To illustrate a non-normal case, Fig.~\ref{fig:non_normal}B shows $\|\delta\bm{x}(t)\|$ from \eqref{eq:delx} with $\alpha<0<\omega$.
As shown in Fig.~\ref{fig:non_normal}B, the perturbation can initially grow no faster than $e^{\omega t}$ even though it eventually decays proportionally to $e^{\alpha t}$.

The imaginary parts of the eigenvalues characterize oscillatory behavior near the fixed point.
In particular, $\nu:=\max\{\Im\lambda\mid\lambda\in\Lambda(\M)\}$ gives the maximum oscillation frequency $\nu/(2\pi)$ around the fixed point.

Quantities such as $\alpha$, $\omega$, and $\nu$ have long been a subject of discussion in random matrix theory (RMT)~\cite{allesina2015stability, gardner1970connectance, may1972will, allesina2012stability, tang2014correlation, tang2014reactivity, mambuca2022dynamical, yang2023reactivity, valigi2024local}. While RMT provides specific expressions for these quantities, there is a problem in that its application to real ecosystems does not reflect reality. 
RMT typically assumes that the elements of $\M$ are random and considers the limit $N\to\infty$.
However, real ecosystems are not necessarily random, and $N$ is often not large.
In fact, it has been reported that RMT fails to explain the behavior of real ecosystems in analyses using real data~\cite{james2015constructing, jacquet2016no}. The motivation for this research is to consider valid constraints on these quantities in systems with finite $N$ and no appropriate randomness that cannot be handled by RMT.

\subsection{Recent studies on the bound on the numerical range}
Recent advances in stochastic thermodynamics have established spectral and numerical-range constraints from a thermodynamic perspective~\cite{barato2017coherence,uhl2019affinity,ohga2023thermodynamic,xu2025thermodynamic, kolchinsky2026cycle}. Unlike RMT, this theory is attracting attention as a constraint that does not assume randomness and holds for finite $N$.
In Ref.~\cite{xu2025thermodynamic}, the authors showed that the numerical range $W(\widetilde{\mathsf{R}})$ of a transformed Markov rate matrix $\widetilde{\mathsf{R}}$ is contained in an ellipse, i.e., $W(\widetilde{\mathsf{R}})\subseteq E(a, b, c)$ with
\begin{align}\label{eq:ellipse}
    E(a, b, c):=\left\{z\in\mathbb{C}\;\middle|\;\left(\frac{\Re z - c}{a}\right)^2 + \left(\frac{\Im z}{b}\right)^2 \le 1\right\}.
\end{align}
Here, $a$ and $-c$ are determined by the maximum escape rate that is given by the diagonal elements of $\widetilde{\mathsf{R}}$. Due to a specific property of the Markov rate matrix and its transform, $a$ can also be related to the symmetric property of $\widetilde{\mathsf{R}}$.
The quantity $b/a$ corresponds to the thermodynamic affinity that quantifies the skew-symmetry of $\widetilde{\mathsf{R}}$.
This result provides a spectral constraint based on thermodynamic quantities. Moreover, this result can yield a tighter bound than the Gershgorin circle theorem ~\cite{horn2012matrix}, which provides general spectral bounds.

Focusing on the mathematical aspects of this result, we extend the result so that it holds for $\M$, thereby deriving general constraints on $\alpha$, $\omega$, and $\nu$ that do not require assumptions about randomness or infiniteness of $N$ for the RMT to hold. As shown later, if there is a certain degree of randomness, this result is complementary to the RMT prediction and can be expressed solely in terms of equivalent quantities used in RMT.

\section{Results}
\subsection{Bounds on the characteristics of linear response}
\begin{figure}[htbp]
    \centering
    \includegraphics[width=\linewidth]{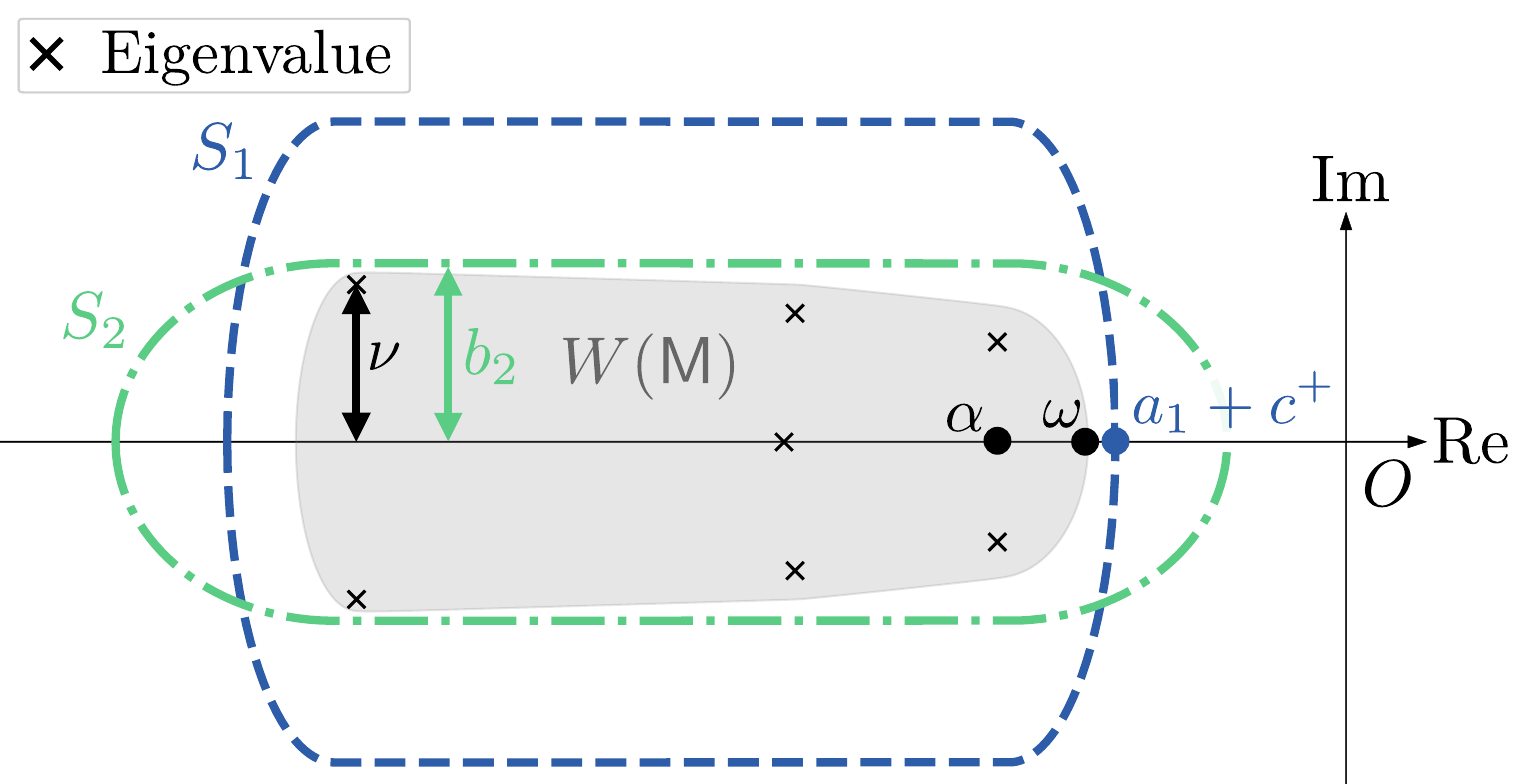}
    \caption{An example of eigenvalues of $\M$ (black crosses), the numerical range $W(\M)$ (gray region), and $S_1$ (blue dashed line) and $S_2$ (green dash-dotted line). The quantity $b_2$ is the upper bound on $\nu$, and $a_1 + c^+$ is an upper bound on both the spectral abscissa $\alpha$ and the numerical abscissa $\omega$. The relation $W(\M)\subseteq S_1\cap S_2$ holds, and the eigenvalues lie in the numerical range $W(\M)$.
    See SI Appendix for details of the numerical calculation used to generate this figure.}
    \label{fig:S1_cap_S2}
\end{figure}

As an extension of Ref.~\cite{xu2025thermodynamic}, our main result provides a bound on $W(\M)$ as the intersection of the ellipses (see Fig.~\ref{fig:S1_cap_S2}; the proof is given in SI Appendix):
\begin{align}\label{eq:bound_W}
    W(\M)\subseteq S_1\cap S_2.
\end{align}
Here, $S_i:=\bigcup_{c \in [c^-, c^+]}E(a_i, b_i, c)$ for $i=1,2$. Unlike in Ref.~\cite{xu2025thermodynamic}, the upper bounds involve two ellipses $E(a_1, b_1, c)$ and $E(a_2, b_2, c)$, and the ellipse center $c$ varies. The quantity $c \in [c^-, c^+]$ is given by the diagonal elements of $\M$ as $c^-:=\min_i (M_{ii})$ and $c^+:=\max_i (M_{ii})$.
The quantity $a_1$ is defined as $a_1:=\min\left\{\left\|\M_+^\mathrm{off}\right\|_\mathrm{F},\,\left\|\M_+^\mathrm{off}\right\|_\infty\right\}$, where $\M_+^\mathrm{off}$ is the off-diagonal part of the symmetrized matrix defined as $(\M_+^\mathrm{off})_{ij}:=(1-\delta_{ij})(\M_+)_{ij}$ with $\M_+:=(\M+\M^\top)/2$. Here, $\left\|\bullet \right\|_\mathrm{F}$ and $\left\|\bullet \right\|_\infty$ denote the Frobenius norm and the induced $\infty$-norm, respectively, i.e., $\left\|\B\right\|_\mathrm{F}:=\sqrt{\sum_{i, j}B_{ij}^2}$ and $\left\|\B\right\|_\infty:=\max_i\sum_j|B_{ij}|$ for a matrix $\B =(B_{ij})$. The quantity $b_1$ is defined through $b_1/a_1$, which represents the degree of skew-symmetry of $\M$, defined by 
$b_1/a_1:=\max_{\substack{i\ne j\mid (M_{ij}, M_{ji})\ne (0, 0)}}|(\M_-)_{ij}/(\M_+)_{ij}|$ with $\M_-:=(\M-\M^\top)/2$. 
Conversely, $b_2$ represents the skew-symmetry of $\M$, and $a_2$ is defined through $a_2/b_2$, which represents the symmetry of $\M$. They are defined as $b_2:=\min\left\{\left\|\M_-\right\|_\mathrm{F},\, \left\|\M_-\right\|_\infty\right\}$ and $a_2/b_2:= \max_{\substack{i\ne j\mid (M_{ij}, M_{ji})\ne (0, 0)}}|(\M_+)_{ij}/(\M_-)_{ij}|$. Since the definitions of these quantities involve division or maximum values, this definition does not apply when a denominator vanishes or the maximization set is empty. Such exceptions must be defined separately using $0$ or $\infty$ (See Materials and Methods).

Note that $-c^\pm$ quantifies the diagonal elements of $\M$,
$a_i$ (or $a_i/b_i$) quantifies the overall symmetry of $\M$, whereas $b_i$ (or $b_i/a_i$) quantifies the overall skew-symmetry, analogous to $-c$, $a$ and $b/a$ in Ref.~\cite{xu2025thermodynamic}.
We also note that \eqref{eq:bound_W} holds for any finite-dimensional real matrix, not only for the Jacobian matrix $\M$ arising from the gLV equations.

The bound \eqref{eq:bound_W} directly yields bounds on $\alpha$, $\omega$, and $\nu$, which determine the behavior of $\|\delta\bm{x}(t)\|$ around the fixed point.
From \eqref{eq:bound_W}, $\omega\le\min\{a_1, a_2\}+c^+$ and $\nu\le \min\{b_1, b_2\}$ follow.
In addition, $a_1\le a_2$ and $b_2\le b_1$ hold, as shown in SI Appendix.
Combining these relations yields the bounds
\begin{gather}
    \alpha\le\omega\le a_1+c^+,\label{eq:bound_omega}\\
    \nu\le b_2,\label{eq:bound_nu}
\end{gather}
(see also Fig.~\ref{fig:S1_cap_S2}).

\begin{figure*}[htbp]
    \centering
    \includegraphics[width=\linewidth]{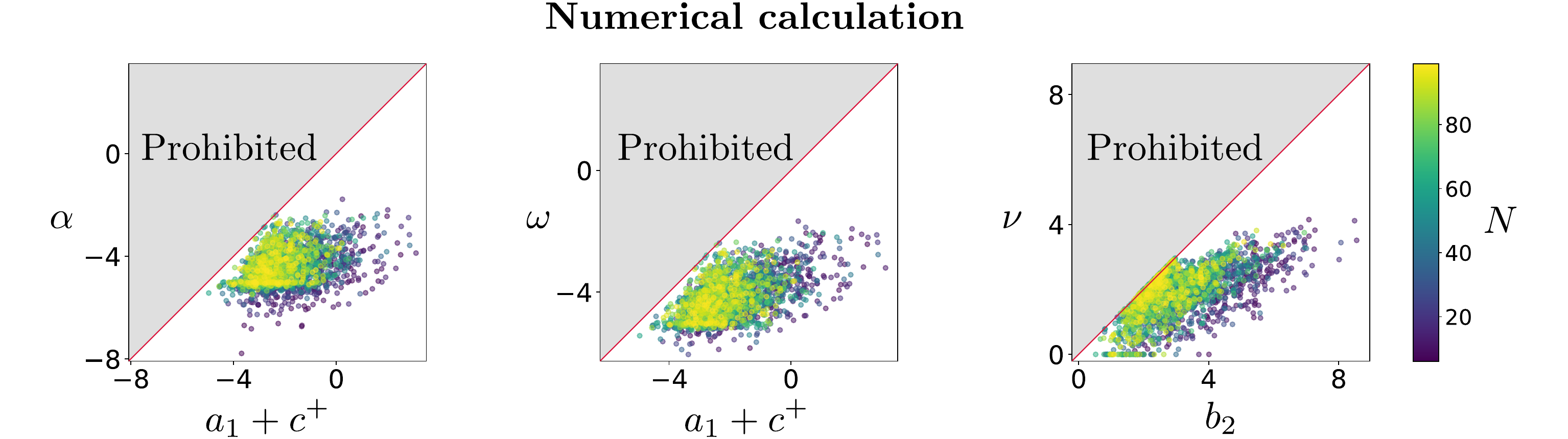}
    \caption{Numerical verification of Eqs.~(\ref{eq:bound_omega}) and (\ref{eq:bound_nu}).
    Plots of $\alpha$ versus $a_1+c^+$, $\omega$ versus $a_1+c^+$, and $\nu$ versus $b_2$ for random matrices $\M$.
    The red solid lines represent $y=x$.
    No points can lie in the region $y>x$ because of the bounds [Eqs.~(\ref{eq:bound_omega}) and (\ref{eq:bound_nu})]. It can also be seen that this holds true for various matrix sizes $N$.
    See SI Appendix for details of the numerical calculation.}
    \label{fig:omega_nu_bound}
\end{figure*}

Ecologically, for $i\ne j$, $(\M)_{ij}=\xbar_iA_{ij}$ is regarded as the interspecific interaction strength weighted by $\xbar_i$, whereas $(\M)_{ii}=\xbar_iA_{ii}$ is the weighted intraspecific competition.
Therefore, $c^+$, $a_1$ and $b_2$ represent the weighted intraspecific competition, the overall symmetry of the weighted interaction strengths, and the overall skew-symmetry of the weighted interaction strengths, respectively. Thus, Eqs.~(\ref{eq:bound_omega}) and (\ref{eq:bound_nu}) show that stability is controlled by the weighted intraspecific competition and the symmetry of the weighted interaction strengths, whereas oscillatory behavior is governed by their skew-symmetry.
These bounds also connect linear response to interaction type.
Cooperative and competitive interactions tend to increase $a_1$, because $(\M)_{ij}$ and $(\M)_{ji}$ have the same sign. Consequently, these interactions can yield large values of $\alpha$ and $\omega$.
In contrast, predator--prey interactions tend to increase $b_2$, because $(\M)_{ij}$ and $(\M)_{ji}$ have opposite signs.
Consequently, predator--prey interactions can yield large values of $\nu$. In Fig.~\ref{fig:omega_nu_bound}, we numerically verify Eqs.~(\ref{eq:bound_omega}) and (\ref{eq:bound_nu}).

\subsection{Alternative bounds}
Since \eqref{eq:bound_W}, from which Eqs.~(\ref{eq:bound_omega}) and (\ref{eq:bound_nu}) follow, holds for general real matrices, it imposes the same constraints on similar matrices with the same eigenvalues. Therefore, in some cases, it may be more ecologically meaningful to consider different similar matrices rather than $\M$. As shown below, it is possible to consider similar matrices to $\M$ that have different ecological interpretations by using a diagonal matrix based on fixed-point values $\X:=\mathrm{diag}(\xbar_1,\ldots,\xbar_N)$.

The first option is to apply the results to a similar matrix $\widetilde{\M}:=\X^{-1}\M\X=\A\X$. For a rescaled perturbation $\delta\widetilde{x}_i(t):=\delta x_i(t)/\xbar_i$ , this matrix describes the behavior around a fixed point as $\frac{d}{dt}\delta\widetilde{\bm{x}}(t)=\widetilde{\M}\delta\widetilde{\bm{x}}(t)$. Since the abundance often varies greatly among species~\cite{fisher1943relation, preston1948commonness, whittaker1965dominance}, the scaling of perturbation is natural. Since similarity transformations preserve the spectrum, $\alpha$ and $\nu$ remain unchanged.
We therefore obtain alternative bounds on $\alpha$ and $\nu$ by applying the same procedure used to derive \eqref{eq:bound_omega} and \eqref{eq:bound_nu} to $\widetilde{\M}$. In the alternative bounds, $\omega$, $a_1$, and $b_2$ are replaced by their counterparts defined with $\widetilde{\M}$ instead of $\M$. See Materials and Methods for explicit expressions for these bounds.

The second option corresponds to a situation where we consider a similar matrix $\widehat{\M}:=\X^{-1/2}\M\X^{1/2}=\X^{1/2}\A\X^{1/2}$ and modify the definition of the rescaled perturbation as $\delta\widehat{x}_i(t):=\delta x_i(t)/\sqrt{\xbar_i}$, which satisfies $\frac{d}{dt}\delta\widehat{\bm{x}}(t)=\widehat{\M}\delta\widehat{\bm{x}}(t)$. As for $\widetilde{\M}$, alternative bounds can be obtained without changing the spectrum. In particular, the bounds associated with $\widehat{\M}$ clarify how interaction structure shapes the linear response.
Since the symmetrization and skew-symmetrization of $\widehat{\M}$ itself are linked to the symmetrization and skew-symmetrization of $\A$ as $(\widehat{\M} \pm  \widehat{\M}^{\top})/2=\X^{1/2}[(\A\pm \A^{\top})/2]\X^{1/2}$, the symmetric and skew-symmetric parts of $\A$ directly affect the bounds. If all off-diagonal interactions are skew-symmetric, the upper bound on $\alpha$ depends only on $c^+$.
Conversely, if all interactions are symmetric, the upper bound on $\nu$ vanishes and hence $\nu=0$ (see Materials and Methods for details).

\subsection{Comparison with previous studies in random matrix theory}
Although our results above do not rely on RMT, it is beneficial to compare them with conventional RMT-based results.
When the matrix elements are random, we can express our results in terms of the same quantities used in RMT-based results.
Nonetheless, our results are valid for any finite $N$, whereas RMT-based ones are valid only in the limit $N\to\infty$.

We first review conventional RMT-based results.
Suppose that we generate $\M$ as a random matrix whose off-diagonal pairs are connected with probability $C$. 
Each connected off-diagonal pair is assigned two correlated, zero-mean random variables with variance $\sigma^2$, correlation coefficient $\rho$, and finite fourth moment. 
Disconnected pairs are set to zero, and all diagonal elements are set to $-d$.
For sufficiently large $N$, the following asymptotic expressions are known to hold~\cite{tang2014reactivity} (see SI Appendix for a precise statement):
\begin{align}
    \alpha&\approx-d+\sigma(1+\rho)\sqrt{NC},\label{eq:alpha_RMT}\\
    \omega&\approx-d+\sigma\sqrt{2(1+\rho)NC},\label{eq:omega_RMT}\\
    \nu&\approx\sigma(1-\rho)\sqrt{NC}\label{eq:nu_RMT}.
\end{align}
In particular, a sufficient condition for $\alpha<0$ (or $\omega<0$) is $\sigma(1+\rho)\sqrt{NC}<d$ (or $\sigma\sqrt{2(1+\rho)NC}<d$).
We notice that Eqs.~(\ref{eq:alpha_RMT})--(\ref{eq:nu_RMT}) can be generalized (see Ref.~\cite{tang2014reactivity} and SI Appendix), but we compare Eqs.~(\ref{eq:alpha_RMT})--(\ref{eq:nu_RMT}) with our results for simplicity.

To compare our results with RMT-based results, we randomize $\M$ as above. Since our bounds hold for each fixed $\M$ and we do not need to consider convergence in the large-$N$ limit, we will use the expected values of $\alpha$, $\omega$, and $\nu$ for the comparison.
Let $\mathbb{E}[\bullet]$ denote the expectation over realizations of $\M$. 
Then, Eqs.~(\ref{eq:bound_omega}) and (\ref{eq:bound_nu}) imply 
\begin{gather}
    \mathbb{E}[\alpha]\le\mathbb{E}[\omega]\le-d+\sigma\sqrt{(1+\rho)\frac{N(N-1)}{2}C},\label{eq:E_omega_bound}\\
    \mathbb{E}[\nu]\le\sigma\sqrt{(1-\rho)\frac{N(N-1)}{2}C},\label{eq:E_nu_bound}
\end{gather}
(see Materials and Methods).
In particular, a sufficient condition for $\mathbb{E}[\omega]<0$ (and hence for $\mathbb{E}[\alpha]<0$) is $\sigma\sqrt{(1+\rho)\frac{N(N-1)}{2}C}<d$. A major difference from RMT is that these inequalities hold for
every finite $N$.

We compare our bounds [Eqs.~(\ref{eq:E_omega_bound}) and (\ref{eq:E_nu_bound})] with their RMT counterparts [Eqs.~(\ref{eq:alpha_RMT})--(\ref{eq:nu_RMT})]. The two results show broadly similar dependence on $d$, $\sigma$ and $C$. Although they differ in their scaling with respect to $\rho$ and $N$, they show qualitatively similar trends. This suggests that upper bounds of Eqs.~(\ref{eq:E_omega_bound}) and (\ref{eq:E_nu_bound}) reflect, at least qualitatively, the large-$N$ asymptotic behavior. However, since the scaling with $N$ differs, our bounds tend to overestimate the corresponding RMT asymptotic behavior as $N$ becomes large, making the inequalities increasingly loose.
This overestimation reflects the fact that Eqs.~(\ref{eq:E_omega_bound}) and (\ref{eq:E_nu_bound}) are derived from Eqs.~(\ref{eq:bound_omega}) and (\ref{eq:bound_nu}), which remain valid even for individual events that become probabilistically negligible in the large-$N$ limit  (see also the precise statement of Eqs.~(\ref{eq:alpha_RMT})--(\ref{eq:nu_RMT}) in SI Appendix).
Our results therefore complement RMT-based ones, particularly for finite $N$, where the behavior cannot be fully characterized by the $N\to\infty$ limit.

\subsection{Real data analysis}
\begin{figure*}[htbp]
    \centering
    \includegraphics[width=0.95\linewidth]{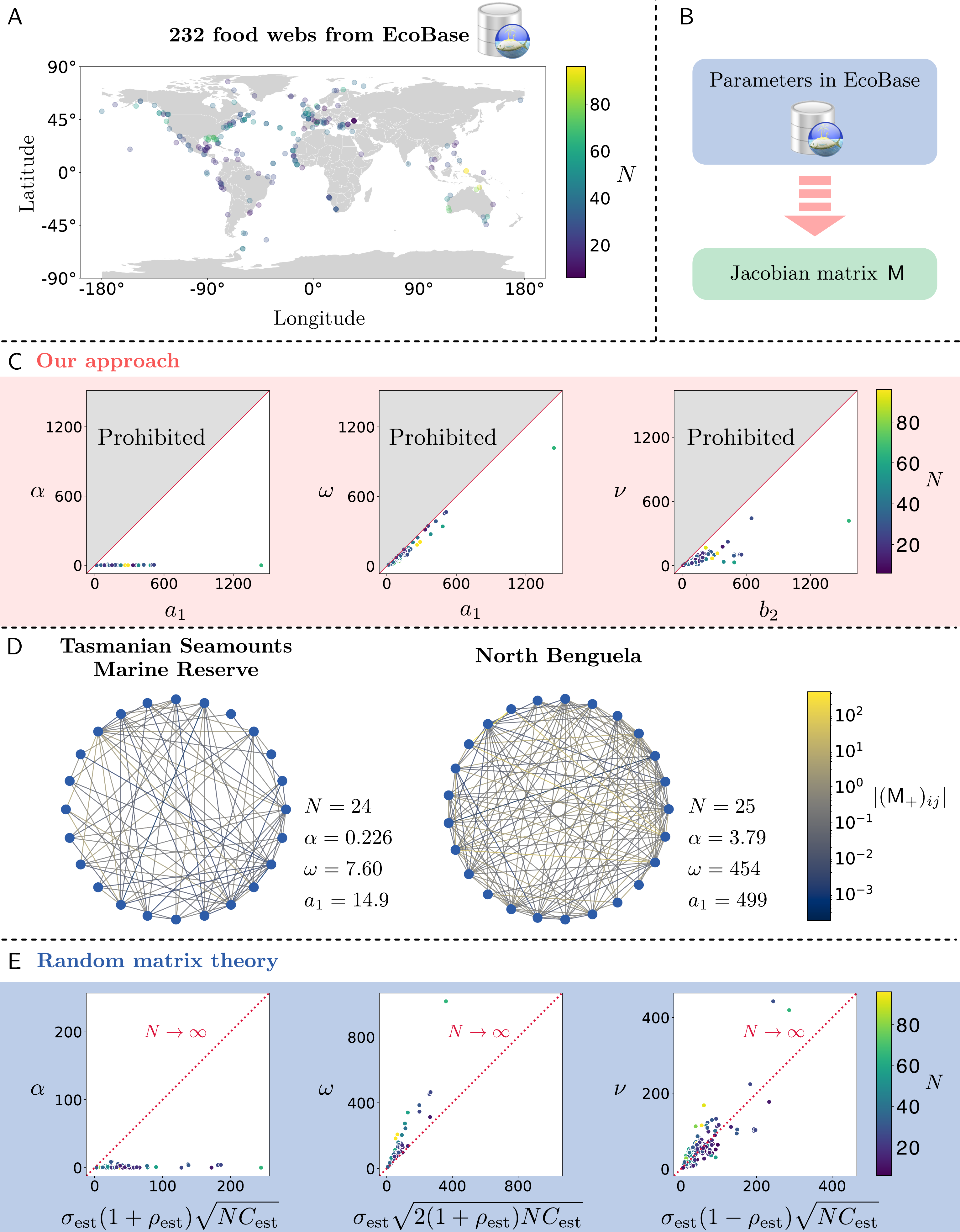}
    \caption{
    (A) Location of the ecosystem data in EcoBase.
    Each circle corresponds to a single ecosystem, and its color represents the number of species $N$ in the ecosystem.
    (B) Schematic of conversion of parameters in EcoBase into the Jacobian matrix $\M$.
    (C) Empirical verification of our results, Eqs.~(\ref{eq:bound_omega}) and (\ref{eq:bound_nu}), using EcoBase.
    The red solid lines represent $y=x$.
    No data points can lie in the region $y>x$, indicating Eqs.~(\ref{eq:bound_omega}) and (\ref{eq:bound_nu}) are valid. 
    (D) Networks of two ecosystems (Tasmanian Seamounts Marine Reserve and North Benguela) in EcoBase with small and large values of $a_1$.
    Each node represents a species, and the color of each edge represents $|(\M_+)_{ij}|$.
    (E) Empirical verification of RMT-based results, Eqs.~(\ref{eq:alpha_RMT})--(\ref{eq:nu_RMT}), using EcoBase.
    The red dotted lines represent $y=x$, corresponding to RMT predictions in $N\to\infty$.
    }
    \label{fig:EcoBase}
\end{figure*}

We apply our results to actual data with a finite value of $N$ to demonstrate their usefulness.
We also assess RMT-based results by using actual data to compare with our results.
To this end, we analyzed 232 real ecosystems from EcoBase, a standard open-access database of marine food webs~\cite{EcoBase, colleter2013ecobase, colleter2015global} (see Fig.~\ref{fig:EcoBase}A; see Materials and Methods for details on EcoBase).
Following previous studies~\cite{de1995energetics, jacquet2016no, feng2025mediating}, we converted the EcoBase parameters into a Jacobian matrix $\M$ (see Fig.~\ref{fig:EcoBase}B and Materials and Methods).
These Jacobian matrices serve to test our results and the RMT-based relations below.

In Fig.~\ref{fig:EcoBase}C, we test Eqs.~(\ref{eq:bound_omega}) and (\ref{eq:bound_nu}).
Note that $c^+=0$ because $M_{ii}=0$ for all $i$ in this data analysis (see also Materials and Methods).
Each point represents one ecosystem, and its color indicates the number of species $N$.
As shown in Fig.~\ref{fig:EcoBase}C, our results hold for all ecosystems, even for small $N$ (see SI Appendix for detailed analysis).
Our bounds provide informative constraints on $\omega$ and $\nu$, indicating that $a_1$ and $b_2$ can account for their typical behavior. For $\alpha$, however, the bound based on $a_1$ is not tight, because $\alpha$ can be substantially smaller than $\omega$, i.e., $\alpha\ll\omega$. Thus, the bound may considerably overestimate $\alpha$. Such a discrepancy between $\alpha$ and $\omega$ can arise when $\M$ lacks favorable structural properties such as normality. Its occurrence in real data reflects the fact that such properties cannot generally be assumed for $\M$. In SI Appendix, we also verify alternative versions of Eqs.~(\ref{eq:bound_omega}) and (\ref{eq:bound_nu}) mentioned above and observe similar behavior.
Therefore, our approach may be applied to real ecosystems to infer their behavior, particularly with regard to short-term stability and oscillatory behavior, from $a_1$ or $b_2$.

We next present and discuss two representative examples in Fig.~\ref{fig:EcoBase}D illustrating how specific network features affect $\omega$, $\alpha$ and $a_1$.
Each node represents a species, and the color of each edge represents the symmetric part of weighted interaction strength, $|(\M_+)_{ij}|$.
Although the two networks have similar values of $N$, their values of $a_1$ differ substantially.
In the Tasmanian Seamounts Marine Reserve, there are relatively few edges and their weights $|(\M_+)_{ij}|$ are generally small, whereas the North Benguela ecosystem has more edges with larger weights.
As a result, $a_1$ is much larger in the North Benguela ecosystem, consistent with its larger values of $\omega$ and $\alpha$, with all three quantities differing on an order-of-magnitude scale. These representative examples therefore suggest that, although $a_1$ does not accurately capture the value of $\alpha$ because the bound is not tight, it may still explain trends in $\alpha$ if $\omega$ and $\alpha$ are related at the order-of-magnitude level. Indeed, a weak but statistically significant positive correlation is observed between $\alpha$ and $a_1$, with a Pearson correlation coefficient of $0.326$ (see SI Appendix for details).

To compare our approach with RMT, we also test the RMT-based relations Eqs.~(\ref{eq:alpha_RMT})--(\ref{eq:nu_RMT}) in Fig.~\ref{fig:EcoBase}E.
See SI Appendix for the definitions of $\sigma_\mathrm{est}$, $\rho_\mathrm{est}$, and $C_\mathrm{est}$, which are estimators of $\sigma$, $\rho$, and $C$. We note that, to the best of our knowledge, these analyses provide the largest-scale empirical test to date of RMT predictions for $\alpha$ and represent the first empirical assessment of RMT predictions for $\omega$ and $\nu$.
As shown in Fig.~\ref{fig:EcoBase}E, the data do not agree well with the RMT predictions in the $N\to\infty$ limit, indicated by the red dotted lines. 
In particular, RMT tends to overestimate $\alpha$ and underestimate $\omega$, indicating that RMT does not capture the typical behavior of $\alpha$ and $\omega$.
Furthermore, although the overall trend in $\nu$ is similar to that in the $N\to\infty$ limit, its values fluctuate, and several outliers are observed (see SI Appendix for detailed analysis). These findings therefore suggest that, in all cases considered, RMT provides a less satisfactory explanation of real ecosystems than our results, which provide rigorous bounds.
Intriguingly, although RMT is expected to become exact in the large-$N$ limit, the data do not always agree with the RMT predictions even for large $N$. Therefore, RMT predictions may be less accurate not only because $N$ is finite, but also because the assumptions about randomness and correlations among the entries of $\M$ may not hold.
In SI Appendix, we test the more general RMT-based results of Ref.~\cite{tang2014reactivity}, which generalize Eqs.~(\ref{eq:alpha_RMT})--(\ref{eq:nu_RMT}), and observe similar behavior.
Therefore, an RMT-based approach may not always be appropriate for real ecosystems.

\section{Conclusion and Discussion}
In this paper, we propose a theory for predicting the linear response of gLV systems to overcome the limitations of RMT. While RMT is valid only for large ecosystems, our approach remains applicable even to small ecosystems. Our results also hold without specific assumptions about randomness and allow us to characterize the linear response of gLV systems solely in terms of structural quantities such as $a_1$ and $b_2$, which quantify the symmetry and skew-symmetry of interspecific interactions, respectively. Moreover, even when the randomness assumptions used in RMT are satisfied, our results provide complementary bounds that hold for any finite $N$ and are expressed in terms of the same parameters used in RMT. These bounds are distinct from the asymptotic predictions of RMT in the $N\to\infty$ limit. This means that our bounds may become loose in the $N\to\infty$ limit. Our results provide rigorous bounds for finite $N$, but cannot capture the asymptotic behavior predicted by RMT in the $N\to\infty$ limit. Therefore, our results should be viewed as complementary to RMT.

We further analyzed data from real ecosystems to test both RMT and our approach. The empirical results showed that the spectral abscissa, the numerical abscissa, and the maximum oscillation frequency can all deviate from the asymptotic predictions of RMT. In particular, the trends in the spectral and numerical abscissas differed substantially from these predictions. These discrepancies may reflect the finite sizes of real ecosystems and the fact that their interaction structures do not necessarily satisfy the randomness assumptions underlying RMT. In contrast, the bounds provided by our approach were satisfied for every ecosystem analyzed. In particular, the bounds on the numerical abscissa and the maximum oscillation frequency were sufficiently tight to capture the behavior of these quantities.  Although the bound for the spectral abscissa is not so tight, a correlation was also observed between the spectral abscissa and its corresponding bound. These findings suggest that our approach may be more effective than the asymptotic RMT predictions considered here for analyzing real ecosystems, which are finite and do not necessarily satisfy specific randomness assumptions.

The finite-$N$ perspective developed here may also open new directions for studying other properties of gLV systems. For example, recent studies have investigated the number and properties of fixed points in random gLV systems, primarily in the large-$N$ limit, using tools from spin glass theory such as dynamical mean-field theory~\cite{galla2018dynamically, roy2019numerical, baron2023breakdown, azaele2024generalized}, the cavity method~\cite{bunin2017ecological}, and the replica method~\cite{biroli2018marginally, altieri2021properties, Ros2023quenched, ros2023generalized}. The present work focuses on the linear response around a given fixed point and derives rigorous bounds valid for any finite $N$. This suggests a broader direction for future research: exploring whether finite-$N$ analyses, in the spirit of the present work, can yield new bounds or other rigorous relations that complement the results obtained using these and other large-$N$ approaches. Such developments could provide additional constraints on a wider range of properties of gLV systems, with potential applications to real ecosystems with finite numbers of species.

Since our bound on the spectral abscissa was not tight in the empirical analysis, we now discuss possible ways to improve it. Our approach is inspired by a recent study in stochastic thermodynamics~\cite{xu2025thermodynamic}, which derives bounds on the spectra of Markov rate matrices governing master equations by constraining their numerical ranges. As in our approach, the spectrum is thus bounded indirectly through the numerical range, and the resulting bound need not be tight. Indeed, other studies in stochastic thermodynamics~\cite{ohga2023thermodynamic, kolchinsky2026cycle} have shown that directly analyzing the eigenvalues of finite-dimensional Markov rate matrices can yield tighter spectral bounds than those obtained in Ref.~\cite{xu2025thermodynamic}. Extending such methods to our setting could lead to tighter bounds on the spectral abscissa and on individual eigenvalues. However, these methods in stochastic thermodynamics rely on the specific structure of Markov rate matrices, and it is not obvious whether they can be generalized to the linear stability analysis of gLV systems considered here.

Because our bounds hold for arbitrary real square matrices of finite dimension, their potential applications extend beyond the linear stability analysis of ecosystems described by the gLV equations. Examples include neural networks~\cite{pachitariu2026critical}, gene regulatory networks~\cite{chen2019gene}, and financial markets~\cite{plerou2002random}, where random-matrix approaches have been used to investigate spectral properties relevant to empirical systems. Applying our bounds to matrices constructed from empirical data or derived from suitable models could provide rigorous spectral constraints at finite $N$, complementing large-$N$ asymptotic predictions.

\section{Materials and Methods}
\subsection{Precise definitions of $E(a_i, b_i, c)$, $b_1$, and $a_2$}
We first give a more precise definition of $E(a_i, b_i, c)$ by defining a few undefined exceptions.
Note that $a_i$ and $b_i$ may take the values $0$ and (as shown below) $\infty$.
In these cases, the inequality $[(\Re z-c)/a_i]^2+(\Im z/b_i)^2\le 1$ appearing in the definition of $E(a_i, b_i, c)$ is not well defined.
To avoid this issue, we adopt the following convention.
For a finite $a_i$, we interpret $E(a_i, 0, c)$ as $|\Re z-c|\le a_i$ and $\Im z=0$, and $E(a_i, \infty, c)$ as $|\Re z-c|\le a_i$, respectively. We similarly define $E(0, b_i, c)$ and $E(\infty, b_i, c)$ by interchanging the roles of $(\Re z-c, a_i)$ and $(\Im z,b_i)$.
We do not need to consider the case $E(\infty, \infty, c)$, because $a_1$ and $b_2$ are always finite by definition.

We next give more precise definitions of $b_1$ and $a_2$ by defining a few undefined exceptions.
The quantity $\max_{\substack{i\ne j\mid (M_{ij}, M_{ji})\ne (0, 0)}}|(\M_-)_{ij}/(\M_+)_{ij}|$ in the definition of $b_1$ is well defined only if there exists at least one pair $i\ne j$ such that $(M_{ij}, M_{ji})\ne (0,0)$ and, for every such pair, $(\M_+)_{ij}\ne 0$.
If this condition fails, we set $b_1:=\infty$.
Similarly, $\max_{\substack{i\ne j\mid (M_{ij}, M_{ji})\ne (0, 0)}}|(\M_+)_{ij}/(\M_-)_{ij}|$ in the definition of $a_2$ is well defined only if there exists at least one pair $i\ne j$ such that $(M_{ij}, M_{ji})\ne (0,0)$ and, for every such pair, $(\M_-)_{ij}\ne 0$.
If this condition fails, we set $a_2:=\infty$.

\subsection{Details of the alternative bounds}
We derive alternative bounds on $\alpha$ and $\nu$.
We first consider the bounds associated with $\widetilde{\M}$.
Applying the same procedure used to derive \eqref{eq:bound_W} to $\widetilde{\M}$, we obtain
$W(\widetilde{\M})\subseteq\widetilde{S}_1\cap\widetilde{S}_2$.
Here, $\widetilde{S}_i$ is defined by replacing $\M$ with $\widetilde{\M}$ in the definitions of $S_i$, $a_i$, $b_i$, and $c^\pm$.
Although the numerical abscissa
$\widetilde{\omega}:=\sup\{\Re \lambda\mid\lambda\in W(\widetilde{\M})\}$
generally differs from $\omega$, $\alpha$ and $\nu$ are unchanged because similarity transformations preserve the spectrum.
Therefore, $W(\widetilde{\M})\subseteq\widetilde{S}_1\cap\widetilde{S}_2$ implies
\begin{gather}
    \alpha\le\widetilde{\omega}\le\widetilde{a}_1+c^+,\\
    \nu\le\widetilde{b}_2.
\end{gather}
Here, $\widetilde{a}_1$ and $\widetilde{b}_2$ are defined by replacing $\M$ with $\widetilde{\M}$ in the definitions of $a_1$ and $b_2$, respectively.
Note that $c^+$ is invariant under this similarity transformation.

Similarly, applying the same procedure to $\widehat{\M}$ yields
\begin{gather}
    \alpha\le\widehat{\omega}\le\widehat{a}_1+c^+,\\
    \nu\le\widehat{b}_2.
\end{gather}
Here, $\widehat{\omega}$, $\widehat{a}_1$, $\widehat{b}_2$, $\widehat{\M}_\pm$, and $\widehat{\M}_+^\mathrm{off}$ are defined by replacing $\M$ with $\widehat{\M}$ in the definitions of $\omega$, $a_1$, $b_2$, $\M_\pm$, and $\M_+^\mathrm{off}$, respectively.
Since $\widehat{\M}_\pm=\X^{1/2}[(\A\pm \A^{\top})/2]\X^{1/2}$, $\widehat{\M}_+^\mathrm{off}$ is the zero matrix if $A_{ij}=-A_{ji}$ for any $i\ne j$.
Thus, $\widehat{a}_1=0$, and the upper bounds on $\alpha$ and $\widehat{\omega}$ depend only on $c^+$.
Conversely, $\widehat{\M}_-$ is the zero matrix if $A_{ij}=A_{ji}$ for any $i\ne j$.
Thus, $\widehat{b}_2=0$, which implies $\nu=0$ because $\nu \geq 0$.

\subsection{Derivation of Eqs.~(\ref{eq:E_omega_bound}) and (\ref{eq:E_nu_bound})}
We first derive \eqref{eq:E_omega_bound}.
Since $M_{ii}=-d$ and $a_1\le\left\|\M_+^\mathrm{off}\right\|_\mathrm{F}$, \eqref{eq:bound_omega} implies $\omega\le-d+\left\|\M_+^\mathrm{off}\right\|_\mathrm{F}$.
Taking expectations of both sides of this inequality and applying Jensen's inequality, we obtain
\begin{align}\label{eq:E_sqrt}
    \mathbb{E}[\omega]&\le-d+\mathbb{E}\left[\sqrt{\sum_{i\ne j}[(\M_+)_{ij}]^2}\right]\nonumber \\
&\le-d+\sqrt{\sum_{i\ne j}\mathbb{E}\left[[(\M_+)_{ij}]^2\right]}.
\end{align}
Note that
\begin{align}\label{eq:E_M_+^2}    
\mathbb{E}\left[[(\M_+)_{ij}]^2\right]&=\frac{1}{4}\left(\mathbb{E}\left[(M_{ij})^2\right]+\mathbb{E}\left[(M_{ji})^2\right]+2\mathbb{E}\left[M_{ij}M_{ji}\right]\right)\notag\\
&=\frac{1}{2}\sigma^2(1+\rho)C,
\end{align}
where we used $\mathbb{E}\left[(M_{ij})^2\right]=\mathbb{E}\left[(M_{ji})^2\right]= C\sigma^2$ and $\mathbb{E}\left[ M_{ij}M_{ji}\right]= C\rho \sigma^2$. Here, the factor $C$ accounts for the connection probability, and disconnected pairs, which occur with probability $1-C$, have $(M_{ij},M_{ji})=(0,0)$ and hence contribute zero to these expectations. 
Substituting \eqref{eq:E_M_+^2} into \eqref{eq:E_sqrt} yields \eqref{eq:E_omega_bound}.

In a similar manner, we prove \eqref{eq:E_nu_bound} as follows. Combining \eqref{eq:bound_nu} with $b_2\le\left\|\M_-\right\|_\mathrm{F}$ and applying Jensen's inequality, as in the derivation of \eqref{eq:E_sqrt}, we obtain
\begin{align}\label{eq:sqrt_E_-}
    \mathbb{E}[\nu]\le\sqrt{\sum_{i\ne j}\mathbb{E}\left[[(\M_-)_{ij}]^2\right]}.
\end{align}
Here, 
\begin{align}\label{eq:E_M_-^2}   
\mathbb{E}\left[((\M_-)_{ij})^2\right]&=\frac{1}{4}\left(\mathbb{E}\left[(M_{ij})^2\right]+\mathbb{E}\left[(M_{ji})^2\right]-2\mathbb{E}\left[M_{ij}M_{ji}\right]\right)\notag\\
&=\frac{1}{2}\sigma^2(1-\rho)C,
\end{align}
and substituting \eqref{eq:E_M_-^2} into \eqref{eq:sqrt_E_-} yields \eqref{eq:E_nu_bound}.

\subsection{Data analysis}
We analyzed EcoBase, an open access database of food webs from around the world~\cite{EcoBase, colleter2013ecobase, colleter2015global}.
See SI Appendix for all data used in this study and their references.
EcoBase is based on the Ecopath with Ecosim (EwE) approach, which is a broadly used approach for modeling food webs~\cite{christensen1992ecopath}.
The EwE approach consists of three main components: (1) Ecopath, a static, mass-balanced snapshot
of the food web; (2) Ecosim, a time-dynamic simulation module; (3) Ecospace, a spatial and temporal dynamic module~\cite{EwE, colleter2013ecobase}.
In this study, we used the Ecopath framework to reconstruct the Jacobian matrices from empirical data.

Following Refs.~\cite{de1995energetics, jacquet2016no, feng2025mediating}, we converted parameters in the Ecopath framework into the Jacobian matrices.
In the Ecopath framework, we can obtain the following parameters: (1) $B_i$, biomass of species $i$; (2) $(P/B)_i$, production/biomass ratio of species $i$; (3) $(Q/B)_i$, consumption/biomass ratio of species $i$; (4) $(DC)_{ij}$, proportion of species $i$ in the diet of species $j$.
Note that not all the parameters were obtained from field sampling; some were instead estimated (see Ref.~\cite{christensen2005ecopath} for details).
We first removed species for which $B_i$, $(P/B)_i$, or $(Q/B)_i$ equals $-9999$, which indicates an invalid entry.
We then converted these parameters into those in the Lotka--Volterra equations as follows.
The biomass of species $i$ was identified with its abundance, i.e., $\xbar_i=B_i$.
If species $i$ is a prey of species $j$ and species $j$ is not a prey of species $i$, we set $A_{ij}=-(Q/B)_j\cdot (DC)_{ij}/B_i$ and $A_{ji}=(P/B)_j\cdot (DC)_{ij}/B_i$.
If species $i$ is a prey of species $j$ and species $j$ is a prey of species $i$ at the same time, we set $A_{ij}=-(Q/B)_j\cdot (DC)_{ij}/B_i+(P/B)_i\cdot (DC)_{ji}/B_j$ and $A_{ji}=-(Q/B)_i\cdot (DC)_{ji}/B_j+(P/B)_j\cdot (DC)_{ij}/B_i$.
Since we cannot reconstruct the self-suppression term $A_{ii}$ from the Ecopath framework, we set $A_{ii}=0$ for all $i$, following Ref.~\cite{jacquet2016no}.
We finally obtained the elements of the Jacobian matrix $\M$ through $M_{ij}=\xbar_iA_{ij}$.
Since $A_{ii}=0$, $M_{ii}=0$ for all $i$.

\subsection{Use of AI}
ChatGPT (GPT-4, GPT-5, and GPT-6 series) was used to assist with language editing and code development. 
The Overleaf AI Assistant was used to assist with language editing.
\\
\section*{Data, Materials, and Software Availability}
The Python code and data used to generate the numerical and ecosystem analysis results presented in this study are publicly available on GitHub at https://github.com/Koya508/glv-numerical-ecobase. The code and data corresponding to version v1.0.0 have been archived on Zenodo at https://doi.org/10.5281/zenodo.23974461.
\\
\section*{Acknowledgments}
K.K. is supported by JSPS KAKENHI Grants No.26KJ1040.
S.I. is supported by JSPS KAKENHI Grants No.23K22412, No.23H00467, and No.25K24775, JST ERATO Grant No.JPMJER2302, and UTEC-UTokyo FSI Research Grant Program.

\bibliography{biblio}

\clearpage
\onecolumngrid
\setcounter{section}{0}
\setcounter{subsection}{0}
\setcounter{subsubsection}{0}
\setcounter{equation}{0}
\setcounter{figure}{0}
\setcounter{table}{0}

\renewcommand{\thesection}{S\arabic{section}}
\renewcommand{\theequation}{S\arabic{equation}}
\renewcommand{\thefigure}{S\arabic{figure}}
\renewcommand{\thetable}{S\arabic{table}}
\renewcommand{\theHsection}{SI.\arabic{section}}
\renewcommand{\theHsubsection}{SI.\arabic{section}.\arabic{subsection}}
\renewcommand{\theHsubsubsection}{SI.\arabic{section}.\arabic{subsection}.\arabic{subsubsection}}
\renewcommand{\theHequation}{SI.\arabic{equation}}
\renewcommand{\theHfigure}{SI.\arabic{figure}}
\renewcommand{\theHtable}{SI.\arabic{table}}

\begin{center}
  {\large\bfseries Supporting Information for\par}
  \medskip
  {\large Linear response in generalized Lotka--Volterra systems with a finite number of species\par}
  \medskip
  Koya Katayama and Sosuke Ito
\end{center}
\bigskip

\begin{SIbibliography}

\section{Derivation of $W(\M)\subseteq S_1\cap S_2$}

We prove $W(\M)\subseteq S_1$ and $W(\M)\subseteq S_2$ separately. 
Before doing so, we derive some relations used below.
Let $z\in W(\M)$ be arbitrary. 
By definition, there exists $\bm{u}\in\mathbb{C}^N$ with $\bm{u}^{\dagger}\bm{u}=1$ such that
\begin{align}
    z
    =\bm{u}^{\dagger}\M\bm{u}
    =\sum_i M_{ii}|u_i|^2+\sum_{i\ne j}M_{ij}u_i^*u_j.
    \label{eq:SI_z}
\end{align}
Defining $c:=\sum_i M_{ii}|u_i|^2$, we have
\begin{align}
    c^-\le c\le c^+,
\end{align}
since $\sum_i|u_i|^2=1$ and $|u_i|^2 \geq 0$. 
Using $\Re(u_j^*u_i)=\Re(u_i^*u_j)$ and $\Im(u_j^*u_i)=-\Im(u_i^*u_j)$, \eqref{eq:SI_z} gives
\begin{align}
    \Re z-c
    &=\sum_{i\ne j}(\M_+)_{ij}\Re(u_i^*u_j),
    \label{eq:SI_Re}\\
    \Im z
    &=\sum_{i\ne j}(\M_-)_{ij}\Im(u_i^*u_j).
    \label{eq:SI_Im}
\end{align}
We derive two norm bounds used below. 
Define
\begin{align}
    Q_+
    :=\sum_{i\ne j}|(\M_+)_{ij}|\cdot|u_i|\cdot|u_j|.
\end{align}
The Cauchy--Schwarz inequality yields
\begin{align}
    Q_+
    &\le
    \left(\sum_{i\ne j}(\M_+)_{ij}^2\right)^{1/2}
    \left(\sum_{i\ne j}|u_i|^2\cdot|u_j|^2\right)^{1/2}
    \le \|\M_+^{\mathrm{off}}\|_{\mathrm F},
    \label{eq:SI_Qplus_F}
\end{align}
where we used $\sum_{i\ne j}|u_i|^2\cdot|u_j|^2\le\left(\sum_i|u_i|^2\right)^2=1$ in the last inequality.
Moreover, $|u_i|\cdot|u_j|\le (|u_i|^2+|u_j|^2)/2$ and $|(\M_+)_{ij}|=|(\M_+)_{ji}|$ imply
\begin{align}
    Q_+
    \le
    \frac12\sum_{i\ne j}|(\M_+)_{ij}|
    \left(|u_i|^2+|u_j|^2\right)=\sum_i|u_i|^2\sum_{j(\ne i)}|(\M_+)_{ij}|
    \le \|\M_+^{\mathrm{off}}\|_{\infty},
    \label{eq:SI_Qplus_inf}
\end{align}
where we used $\sum_i|u_i|^2=1$ and the definition of the induced $\infty$-norm in the last inequality.
Combining \eqref{eq:SI_Qplus_F} and \eqref{eq:SI_Qplus_inf}, we obtain
\begin{align}
    Q_+\le a_1.
    \label{eq:SI_Qplus}
\end{align}
For the  quantity \begin{align}
    Q_-:=\sum_{i\ne j}|(\M_-)_{ij}|\cdot|u_i|\cdot|u_j|,
\end{align}
the same argument, using $|(\M_-)_{ij}|=|(\M_-)_{ji}|$, gives
\begin{align}
    Q_-\leq \left(\sum_{i\ne j}(\M_-)_{ij}^2\right)^{1/2}
    \left(\sum_{i\ne j}|u_i|^2\cdot|u_j|^2\right)^{1/2} \leq \|\M_-\|_{\mathrm F},
\end{align}
and
\begin{align}
    Q_-\le
    \frac12\sum_{i\ne j}|(\M_-)_{ij}|
    \left(|u_i|^2+|u_j|^2\right)=\sum_i|u_i|^2\sum_{j(\ne i)}|(\M_-)_{ij}|
    \le \|\M_-\|_{\infty}.
\end{align}
Therefore, we also obtain
\begin{align}
    Q_-\le b_2.
    \label{eq:SI_Qminus}
\end{align}

\paragraph*{Proof of $W(\M)\subseteq S_1$.}
We consider the following two mutually exclusive and exhaustive cases: (I) there exists at least one pair $i\ne j$ such that $(M_{ij}, M_{ji})\ne (0, 0)$, and $(\M_+)_{ij}\ne 0$ for every such pair; (II) otherwise.

We first consider the case (I), where $0<a_1<\infty$ and $0\le b_1<\infty$. 
In this case,
\begin{align}
    r_1
    :=\max_{i\ne j|(M_{ij},M_{ji})\ne(0,0)}
    \left|\frac{(\M_-)_{ij}}{(\M_+)_{ij}}\right|
\end{align}
is well-defined, and $b_1=r_1a_1$. 
\eqref{eq:SI_Re} implies
\begin{align}
    |\Re z-c|
    &\le \sum_{i\ne j}|(\M_+)_{ij}|\cdot|\Re(u_i^*u_j)|.
    \label{eq:SI_Re_bound_plus}
\end{align}
\eqref{eq:SI_Im} implies 
\begin{align}\label{eq:SI_Im_bound_plus}
    |\Im z|&=\left|\sum_{i\ne j | (\M_-)_{ij}\ne 0}(\M_-)_{ij}\Im(u^*_iu_j)\right|\notag\\
    &=\left|\sum_{i\ne j| (\M_+)_{ij}\ne 0}\left(\frac{(\M_-)_{ij}}{(\M_+)_{ij}}\right)(\M_+)_{ij}\Im(u^*_iu_j)\right|\notag\\
    &\le\max_{k\ne l| (\M_+)_{kl}\ne 0}\left|\frac{(\M_-)_{kl}}{(\M_+)_{kl}}\right|\cdot\sum_{i\ne j|(\M_+)_{ij}\ne 0}|(\M_+)_{ij}|\cdot|\Im(u^*_iu_j)|\notag\\
    &=\max_{k\ne l| (M_{kl}, M_{lk})\ne (0, 0)}\left|\frac{(\M_-)_{kl}}{(\M_+)_{kl}}\right|\cdot\sum_{i\ne j| (M_{ij}, M_{ji})\ne (0, 0)}|(\M_+)_{ij}|\cdot|\Im(u^*_iu_j)|\notag\\
    &=r_1\sum_{i\ne j}|(\M_+)_{ij}|\cdot|\Im(u^*_iu_j)|.
\end{align}
In the second line, we used the fact that $(\M_-)_{ij}\ne 0$ implies $(\M_+)_{ij}\ne 0$ in the case (I).
In the fourth line, we used the fact that $(\M_+)_{ij}\ne 0$ is equivalent to $(M_{ij}, M_{ji})\ne (0, 0)$ in the case (I).
In the last line, we used the fact that a pair $(i, j)$ such that $(M_{ij}, M_{ji})=(0, 0)$ does not contribute to the sum.
Therefore, \eqref{eq:SI_Re_bound_plus} and \eqref{eq:SI_Im_bound_plus} give
\begin{align}\label{eq:b1Re_a1Im}
    [b_1(\Re z - c)]^2 + (a_1\Im z)^2&\le\left(b_1\sum_{i\ne j}|(\M_+)_{ij}|\cdot|\Re(u^*_iu_j)|\right)^2+\left(b_1\sum_{i\ne j}|(\M_+)_{ij}|\cdot|\Im(u^*_iu_j)|\right)^2\notag\\
    &=b_1^2\sum_{i\ne j, k\ne l}|(\M_+)_{ij}|\cdot|(\M_+)_{kl}|\left(|\Re(u^*_iu_j)|\cdot|\Re(u^*_ku_l)|+|\Im(u^*_iu_j)|\cdot|\Im(u^*_ku_l)|\right).
\end{align}
Applying the Cauchy--Schwarz inequality, the term in the summand is bounded as
\begin{align}    
|\Re(u^*_iu_j)|\cdot|\Re(u^*_ku_l)|+|\Im(u^*_iu_j)|\cdot|\Im(u^*_ku_l)|
&\le\sqrt{|\Re(u^*_iu_j)|^2+|\Im(u^*_iu_j)|^2}\sqrt{|\Re(u^*_ku_l)|^2+|\Im(u^*_ku_l)|^2}\notag\\
&=|u_i|\cdot|u_j|\cdot|u_k|\cdot|u_l|.
\end{align}
Combining this bound with \eqref{eq:b1Re_a1Im}, and then using \eqref{eq:SI_Qplus}, we have
\begin{align}\label{eq:SI_S1_ellipse}
    \left[b_1(\Re z-c)\right]^2
    +\left(a_1\Im z\right)^2
    \le b_1^2Q_+^2
    \le (a_1b_1)^2.
\end{align}
If $b_1>0$, \eqref{eq:SI_S1_ellipse} implies $z\in E(a_1,b_1,c)\subseteq S_1$. 
If $b_1=0$, \eqref{eq:SI_S1_ellipse} gives $\Im z=0$, while \eqref{eq:SI_Re_bound_plus}, \eqref{eq:SI_Qplus} and $|\Re(u_i^*u_j)| \leq |u_i|\cdot|u_j|$ give $|\Re z-c|\le a_1$. 
Hence $z\in E(a_1,0,c)\subseteq S_1$.

We next consider the case (II).
By definition, $b_1=\infty$ and $S_1=\bigcup_{c \in [c^-, c^+]}\{z\in\mathbb{C} \mid|\Re z-c|\le a_1\}$.
Similarly, \eqref{eq:SI_Re_bound_plus}, \eqref{eq:SI_Qplus} and $|\Re(u_i^*u_j)| \leq |u_i|\cdot|u_j|$ give $|\Re z-c|\le a_1$, and therefore $z\in S_1$. Thus, $W(\M)\subseteq S_1$ in both cases.

\paragraph*{Proof of $W(\M)\subseteq S_2$.}
We again consider two cases: (I)' there exists at least one pair $i\ne j$ such that $(M_{ij}, M_{ji})\ne (0, 0)$, and $(\M_-)_{ij}\ne 0$ for every such pair, (II)' otherwise.

We first consider the case (I)', where $0<b_2<\infty$ and $0\le a_2<\infty$. 
In this case,
\begin{align}
    r_2
    :=\max_{i\ne j|(M_{ij},M_{ji})\ne(0,0)}
    \left|\frac{(\M_+)_{ij}}{(\M_-)_{ij}}\right|
\end{align}
is well-defined, and $a_2=r_2b_2$. 
Interchanging $\M_+$ with $\M_-$, $|\Re z-c|$ with $|\Im z|$, and $\Re(u^*_i u_j)$ with $\Im(u^*_i u_j)$ in the derivations of Eqs. (\ref{eq:SI_Im_bound_plus})  and (\ref{eq:SI_Re_bound_plus}), respectively, we obtain
\begin{align}
    |\Re z-c|
    &\le r_2\sum_{i\ne j}|(\M_-)_{ij}|\cdot|\Re(u_i^*u_j)|,
    \label{eq:SI_Re_bound_minus}\\
    |\Im z|
    &\le \sum_{i\ne j}|(\M_-)_{ij}|\cdot|\Im(u_i^*u_j)|.
    \label{eq:SI_Im_bound_minus}
\end{align}
Therefore, as in \eqref{eq:SI_S1_ellipse}, we obtain
\begin{align}\label{eq:SI_S2_ellipse}
    \left[b_2(\Re z-c)\right]^2
    +\left(a_2\Im z\right)^2
    \le a_2^2Q_-^2
    \le (a_2b_2)^2,
\end{align}
where we used \eqref{eq:SI_Qminus} in the last inequality.
If $a_2>0$, \eqref{eq:SI_S2_ellipse} implies $z\in E(a_2,b_2,c)\subseteq S_2$. 
If $a_2=0$, \eqref{eq:SI_S2_ellipse} gives $\Re z=c$, while \eqref{eq:SI_Im_bound_minus}, \eqref{eq:SI_Qminus} and $|\Im(u_i^*u_j)| \leq |u_i|\cdot|u_j|$ give $|\Im z|\le b_2$. Hence $z\in E(0,b_2,c)\subseteq S_2$.

We next consider the case (II)'.
By definition, $a_2=\infty$ and $S_2=\{z\in\mathbb{C} \mid|\Im z|\le b_2\}$.
Similarly, \eqref{eq:SI_Im_bound_minus}, \eqref{eq:SI_Qminus} and $|\Im(u_i^*u_j)| \leq |u_i|\cdot|u_j|$ give $|\Im z|\le b_2$, and therefore $z\in S_2$. Thus, $W(\M)\subseteq S_2$ in both cases. 

Combining the two inclusions proves $W(\M)\subseteq S_1\cap S_2$.

\section{Derivation of $a_1\le a_2$ and $b_2\le b_1$}

We first prove $b_2\le b_1$. If the condition defining a finite $b_1$ fails, then $b_1=\infty$ and the inequality is immediate. Otherwise,
\begin{align}
    r_1=\max_{i\ne j|(M_{ij},M_{ji})\ne(0,0)}
    \left|\frac{(\M_-)_{ij}}{(\M_+)_{ij}}\right|
\end{align}
is well-defined, and $|(\M_-)_{ij}|\le r_1|(\M_+)_{ij}|$ for every $i\ne j$.
Hence
\begin{align}
    \|\M_-\|_{\mathrm F}
    = \sqrt{ \sum_{i\neq j}|(\M_-)_{ij}|^2}&\le r_1 \sqrt{ \sum_{i\neq j}|(\M_+)_{ij}|^2}= r_1\|\M_+^{\mathrm{off}}\|_{\mathrm F},\\
    \|\M_-\|_{\infty}= \max_i  \sum_{j (\neq i)}|(\M_-)_{ij}|
    &\le r_1 \max_i  \sum_{j (\neq i)}|(\M_+)_{ij}|=r_1\|\M_+^{\mathrm{off}}\|_{\infty}.
\end{align}
Taking the minimum of the two inequalities gives
\begin{align}
    b_2
    =\min\left\{\|\M_-\|_{\mathrm F},\|\M_-\|_{\infty}\right\}
    \le \min\left\{r_1 \|\M_+^{\mathrm{off}}\|_{\mathrm F},r_1 \|\M_+^{\mathrm{off}}\|_{\infty}\right\}
    = r_1a_1
    =b_1.
\end{align}

The proof of $a_1\le a_2$ is analogous. If the condition defining a finite $a_2$ fails, then $a_2=\infty$ and the inequality is immediate. 
Otherwise,
\begin{align}
    r_2=\max_{i\ne j|(M_{ij},M_{ji})\ne(0,0)}
    \left|\frac{(\M_+)_{ij}}{(\M_-)_{ij}}\right|
\end{align}
is well-defined, and $|(\M_+)_{ij}|\le r_2|(\M_-)_{ij}|$ for every $i\ne j$. Interchanging the roles of $\M^{\mathrm{off}}_+$ and $\M_-$, we obtain
\begin{align}
    \|\M_+^{\mathrm{off}}\|_{\mathrm F}
    &\le r_2\|\M_-\|_{\mathrm F},\\
    \|\M_+^{\mathrm{off}}\|_{\infty}
    &\le r_2\|\M_-\|_{\infty}.
\end{align}
Therefore,
\begin{align}
    a_1= \min\left\{ \|\M_+^{\mathrm{off}}\|_{\mathrm F}, \|\M_+^{\mathrm{off}}\|_{\infty}\right\}
    \le \min\left\{ r_2 \|\M_-\|_{\mathrm F}, r_2\|\M_-\|_{\infty}\right\}=  r_2b_2
    =a_2.
\end{align}

\section{Random matrix theory (RMT)}
We show the precise statement of Eqs. (7), (8), and (9) in the main text:
\begin{align}
    \alpha&\approx-d+\sigma(1+\rho)\sqrt{NC},\tag{7}\\
    \omega&\approx-d+\sigma\sqrt{2(1+\rho)NC},\tag{8}\\
    \nu&\approx\sigma(1-\rho)\sqrt{NC}.\tag{9}
\end{align}
We also show more general results based on RMT.

\subsection{Precise statement of Eqs.~(7)--(9)}
To state the precise meaning of Eqs.(7)--(9), we describe how to generate a random matrix $\M_N\in\mathbb{R}^{N\times N}$ (we specified the size $N$ of the matrix).
Similar discussions can be found in Refs.~\cite{allesina2012stability, tang2014correlation, tang2014reactivity}.
Fix $0<C\le1$, $\sigma>0$, $-1\le\rho\le1$, and $d\in\mathbb{R}$ independently of $N$. 
Let $(\xi_1,\xi_2)$ be a real random vector satisfying
\begin{align}
    \mathbb{E}[\xi_1]=\mathbb{E}[\xi_2]=0,\qquad \mathbb{E}[\xi_1^2]=\mathbb{E}[\xi_2^2]=\sigma^2,\qquad \mathbb{E}[\xi_1\xi_2]=\rho\sigma^2,\qquad \mathbb{E}[|\xi_1|^4]+\mathbb{E}[|\xi_2|^4]<\infty.
\end{align}
To construct $\M_N$, we first construct an infinite random matrix $\M_\infty$ as follows.
(i) For each $i>j$, $((\M_\infty)_{ij}, (\M_\infty)_{ji})$ is sampled from the distribution of $(\xi_1,\xi_2)$ with probability $C/2$, and from the distribution of $(\xi_2, \xi_1)$ with probability $C/2$.
(ii) For each $i>j$, $((\M_\infty)_{ij}, (\M_\infty)_{ji})$ is set to $(0, 0)$ with probability $1-C$.
(iii) For each $i$, $(\M_\infty)_{ii}$ is independently sampled from a distribution with mean $-d$ and finite variance.
For each $N\ge1$, we then define $\M_N$ as the matrix formed by the first $N$ rows and first $N$ columns of this infinite matrix $\M_\infty$.
Note that the condition (iii) is looser than that imposed in the main text.
Nonetheless, Eqs.~(7)--(9) hold under these looser conditions.

We state one of the theorems in RMT, \textit{the elliptic law}, to derive Eqs.(7)--(9).
Define
\begin{align}\label{eq:normalize}
    \M_N^\mathrm{norm}:=\frac{\M_N+d\II_N}{\sqrt{\sigma^2NC}}
\end{align}
and its \textit{empirical spectral measure}
\begin{align}
    m_N:=\frac{1}{N}\sum_{i}\delta_{\lambda_i\left(\M_N^\mathrm{norm}\right)},
\end{align}
where $\II_N$ is the identity matrix with size $N$, $\lambda_i\left(\M_N^\mathrm{norm}\right)$ is the eigenvalue of the normalized matrix $\M_N^\mathrm{norm}$, and $\delta_{\lambda_i\left(\M_N^\mathrm{norm}\right)}$ is the Dirac measure.
Let $\{\M_N^\mathrm{norm}\}_{N\ge 1}$ be a sequence of random matrices generated according to the above procedure and \eqref{eq:normalize}.
The elliptic law (Theorem 1.5 in Ref.~\cite{Nguyen2015}) states that, for $-1<\rho<1$, $m_N$ converges almost surely to $\mathrm{Unif}(E_\rho)$, i.e.,
\begin{align}
    m_N\longrightarrow\mathrm{Unif}(E_\rho).
\end{align}
Here, $\mathrm{Unif}(E_\rho)$ denotes the uniform probability measure on the ellipse $E_\rho$ defined by
\begin{align}
    E_\rho:=\Biggl\{z\in\mathbb{C} \Biggm|\left(\frac{\Re z}{1+\rho}\right)^2 + \left(\frac{\Im z}{1-\rho}\right)^2 \le 1\Biggr\}
\end{align}
for $-1<\rho<1$.
For $\rho=\pm1$, Wigner's semicircle law (Theorem 1.1 in Ref.~\cite{Nguyen2015}) and its analogue on the imaginary axis imply that $m_N$ converges almost surely to a nonuniform distribution supported on $E_1:=\{x\in\mathbb{R}\mid -2\le x\le 2\}$ for $\rho=1$ and on $E_{-1}:=\{iy\in\mathbb{C}\mid -2\le y\le 2\}$ for $\rho=-1$.
Note that the elliptic law and Wigner's semicircle law are not sufficient for deriving Eqs.~(7)--(9).
Indeed, $\sup_{z\in E_\rho}\operatorname{Re}z=1+\rho$ and $\sup_{z\in E_\rho}\operatorname{Im}z=1-\rho$ need not coincide with the spectral abscissa and the maximum oscillation frequency of $\M_N^\mathrm{norm}$, respectively.
This is because a small number of eigenvalues outside $E_\rho$ can determine the spectral abscissa and the maximum oscillation frequency, while having a negligible effect on the empirical spectral measure $m_N$.
Thus, Eqs.~(7) and (9) do not follow from these theorems alone.

Nonetheless, combining these theorems with a result in Ref.~\cite{orourke2014low}, we can derive  Eqs.~(7)--(9) as follows.
Theorem 2.2 in Ref.~\cite{orourke2014low} states that, almost surely, the maximum distance from the eigenvalues of $\M_N^\mathrm{norm}$ to $E_\rho$ becomes arbitrarily small as $N\to\infty$.
This theorem therefore rules out outlier eigenvalues.
This theorem, together with the elliptic law and Wigner's semicircle law, leads to the asymptotic behavior of the spectral abscissa $\alpha_N$ and the maximum oscillation frequency $\nu_N$ of $\M_N$.
That is, the following holds almost surely:
\begin{align}
    \frac{\alpha_N+d}{\sigma\sqrt{NC}}&\longrightarrow 1+\rho,\label{eq:alpha_limit}\\
    \frac{\nu_N}{\sigma\sqrt{NC}}&\longrightarrow 1-\rho.
\end{align}
These limits are the precise meaning of Eqs.~(7) and (9), respectively.
Furthermore, note that the numerical abscissa $\omega_N$ can be written as $\omega_N=\max_i\lambda_i((\M_N)_+)$ with $(\M_N)_+:=(\M_N+\M_N^\top)/2$~\cite{trefethen2020spectra}, and the real random vector associated with this symmetrization $(\xi_{+, 1}, \xi_{+, 2}):=((\xi_1+\xi_2)/2, (\xi_1+\xi_2)/2)$ satisfies
\begin{align}
    \mathbb{E}[\xi_{+, 1}]
=
\mathbb{E}[\xi_{+, 2}]
=0,
\quad
\mathbb{E}[(\xi_{+, 1})^2]
=
\mathbb{E}[(\xi_{+, 2})^2]
=\frac{1}{2}\sigma^2(1+\rho)=:\sigma_+^2,
\quad
\mathbb{E}[\xi_{+, 1}\xi_{+, 2}]
=\frac{1}{2}\sigma^2(1+\rho)=:\rho_+\sigma_+^2,
\end{align}
and
\begin{align}
    \mathbb{E}[|\xi_{+, 1}|^4]+\mathbb{E}[|\xi_{+, 2}|^4]<\infty.
\end{align}
Here, $\rho_+=1$.
Therefore, when $\rho\ne -1$ and thus $\sigma_+\ne 0$, by following a parallel discussion deriving \eqref{eq:alpha_limit}, one finds that the following holds almost surely:
\begin{align}
    \frac{\omega_N+d}{\sigma_+\sqrt{NC}}&\longrightarrow 1+\rho_+=2.
\end{align}
This limit is the precise meaning of Eq.~(8) for $\rho\ne -1$.
When $\rho=-1$, the off-diagonal part of $\M_N$ is skew-symmetric,
so $(\M_N)_+$ is diagonal and
$\omega_N=\max_i(\M_N)_{ii}$.
Since the diagonal elements are random variables with mean $-d$ and finite variance, $(\omega_N+d)/\sqrt{N}\to 0$ almost surely, which is the precise meaning of Eq.~(8) for $\rho=-1$.

\subsection{More general result}
We show more general results compared to Eqs.(7)--(9) in the main text.
See Refs.~\cite{tang2014correlation, tang2014reactivity} for the derivation.
Fix $\mu^\mathrm{all}\in\mathbb{R}$, $\sigma^\mathrm{all}>0$, $-1\le\rho^\mathrm{all}\le1$, and $d\in\mathbb{R}$ independently of $N$. 
Let $(\eta_1,\eta_2)$ be a real random vector satisfying
\begin{align}
    \mathbb{E}[\eta_1]=\mathbb{E}[\eta_2]=\mu^\mathrm{all},\quad \mathbb{E}[\eta_1^2]=\mathbb{E}[\eta_2^2]=\left(\sigma^\mathrm{all}\right)^2+\left(\mu^\mathrm{all}\right)^2,\quad \mathbb{E}[\eta_1\eta_2]-\mathbb{E}[\eta_1]\mathbb{E}[\eta_2]=\rho^\mathrm{all}\left(\sigma^\mathrm{all}\right)^2,\quad \mathbb{E}[|\eta_1|^4]+\mathbb{E}[|\eta_2|^4]<\infty.
\end{align}
To construct $\M_N\in\mathbb{R}^{N\times N}$, we construct an infinite random matrix $\M_\infty$ as follows.
(i)' For each $i>j$, $((\M_\infty)_{ij}, (\M_\infty)_{ji})$ is independently sampled from the distribution of $(\eta_1, \eta_2)$.
(ii)' For each $i$, $(\M_\infty)_{ii}$ is independently sampled from a distribution with mean $-d$ and finite variance.
For each $N\ge1$, $\M_N$ is then defined as the matrix formed by the first $N$ rows and first $N$ columns of $\M_\infty$.
Then, the following holds for large $N$:
\begin{align}
    \alpha&\approx-d-\mu^\mathrm{all}+\max\left\{N\mu^\mathrm{all},\, \sigma^\mathrm{all}(1+\rho^\mathrm{all})\sqrt{N}\right\},\label{eq:alpha_RMT_general}\\
    \omega&\approx-d-\mu^\mathrm{all}+\max\left\{N\mu^\mathrm{all},\, \sigma^\mathrm{all}\sqrt{2(1+\rho^\mathrm{all})N}\right\},\label{eq:omega_RMT_general}\\
    \nu&\approx\sigma^\mathrm{all}(1-\rho^\mathrm{all})\sqrt{N}.\label{eq:nu_RMT_general}
\end{align}
These relations are generalizations of Eqs.(7)--(9).
Indeed, by setting $\mu^\mathrm{all}=0$, $\sigma^\mathrm{all}=\sqrt{C}\sigma$, and $\rho^\mathrm{all}=\rho$, the conditions (i)' and (ii)' are equivalent to the conditions (i)--(iii) in the previous subsection.
Furthermore, \eqref{eq:alpha_RMT_general}, \eqref{eq:omega_RMT_general}, and \eqref{eq:nu_RMT_general} reduce to Eq.~(7), Eq.~(8), and Eq.~(9), respectively.

\section{Details of the empirical verification}
We provide details of the empirical verification of our approach [Eqs.~(5) and (6), and Eqs.~(12)--(15)] and of the RMT-based predictions [Eqs.~(7)--(9) and Eqs.~(\ref{eq:alpha_RMT_general})--(\ref{eq:nu_RMT_general})].

\subsection{Definitions of the estimators used in Fig.~4E}
To test Eqs.~(7)--(9) empirically, we estimated the parameters appearing in these equations as follows.
Defining $\mathcal{E}_\mathrm{nz}:=\{(i, j)\mid i\ne j, (M_{ij}, M_{ji})\ne(0, 0)\}$ and $\mathcal{E}_\mathrm{nz}^>:=\{(i, j)\mid i> j, (M_{ij}, M_{ji})\ne(0, 0)\}$, the estimators of $\sigma^2$, $\rho$, and $C$ are
\begin{gather}
    (\sigma_\mathrm{est})^2:=\frac{1}{|\mathcal{E}_\mathrm{nz}|}\sum_{(i, j)\in\mathcal{E}_\mathrm{nz}}(M_{ij})^2-\left(\frac{1}{|\mathcal{E}_\mathrm{nz}|}\sum_{(i, j)\in\mathcal{E}_\mathrm{nz}}M_{ij}\right)^2,\\
    \rho_\mathrm{est}:=\frac{1}{\left(\sigma_\mathrm{est}\right)^2}\left[\frac{1}{|\mathcal{E}_\mathrm{nz}^>|}\sum_{(i, j)\in\mathcal{E}_\mathrm{nz}^>}M_{ij}M_{ji}-\left(\frac{1}{|\mathcal{E}_\mathrm{nz}|}\sum_{(i, j)\in\mathcal{E}_\mathrm{nz}}M_{ij}\right)^2\right],\\
    C_\mathrm{est}:=\frac{|\mathcal{E}_\mathrm{nz}^>|}{N(N-1)/2}.
\end{gather}
Here, $|\mathcal{E}_\mathrm{nz}|$ and $|\mathcal{E}_\mathrm{nz}^>|$ denote the numbers of elements in $\mathcal{E}_\mathrm{nz}$ and $\mathcal{E}_\mathrm{nz}^>$, respectively.

\subsection{Detailed analysis of Figs.~4C and 4E}
We provide a more detailed analysis of our approach [Eqs.~(5) and (6)] in Fig.~4C and of RMT [Eqs.~(7)--(9)] in Fig.~4E.
We first evaluated the accuracy of each method.
For each point in Figs.~4C and 4E, we computed the ratio of the y-axis value to the x-axis value, $y/x$.
This ratio equals $1$ when the prediction is accurate and is smaller (larger) than $1$ when the method overestimates (underestimates) the quantity.
Figure~\ref{fig:histograms} shows histograms of these ratios.
As shown in Fig.~\ref{fig:histograms}A, the ratio for our approach is always smaller than $1$ (the black dotted line marks $1$), indicating that the bounds in Eqs.~(5) and (6) hold for all ecosystems.
In contrast, for RMT the ratio can be smaller or larger than $1$ (Fig.~\ref{fig:histograms}B).
In particular, the ratio is always smaller than $1$ for $\alpha$, whereas it is typically larger than $1$ for $\omega$.
Thus, for real ecosystems, RMT tends to overestimate $\alpha$ and underestimate $\omega$. This variation in the direction of bias across quantities may be a drawback of RMT-based predictions compared with our rigorous bounds.

We also assessed how well our approach and RMT capture qualitative trends in $\alpha$, $\omega$, and $\nu$ across ecosystems, rather than their exact values.
To this end, we computed Pearson correlation coefficients for each plot in Figs.~4C and 4E.
For our approach (Fig.~4C), the Pearson correlation coefficients were $0.326$, $0.984$, and $0.815$ for $\alpha$, $\omega$, and $\nu$, respectively.
For RMT (Fig.~4E), the Pearson correlation coefficients were $0.385$, $0.949$, and $0.818$ for $\alpha$, $\omega$, and $\nu$, respectively.
For all six plots, p-values were smaller than $10^{-3}$ under the null hypothesis that the variables are uncorrelated and normally distributed.
These results indicate that both our approach and RMT capture the overall tendencies of $\alpha$, $\omega$, and $\nu$ in real ecosystems, because all Pearson correlation coefficients are positive and statistically significant.
Both methods capture the trend particularly well for $\omega$ and $\nu$ (large Pearson correlation coefficients) but less well for $\alpha$ (smaller Pearson correlation coefficients).

\clearpage

\begin{figure}[htbp]
    \centering
    \includegraphics[width=\linewidth]{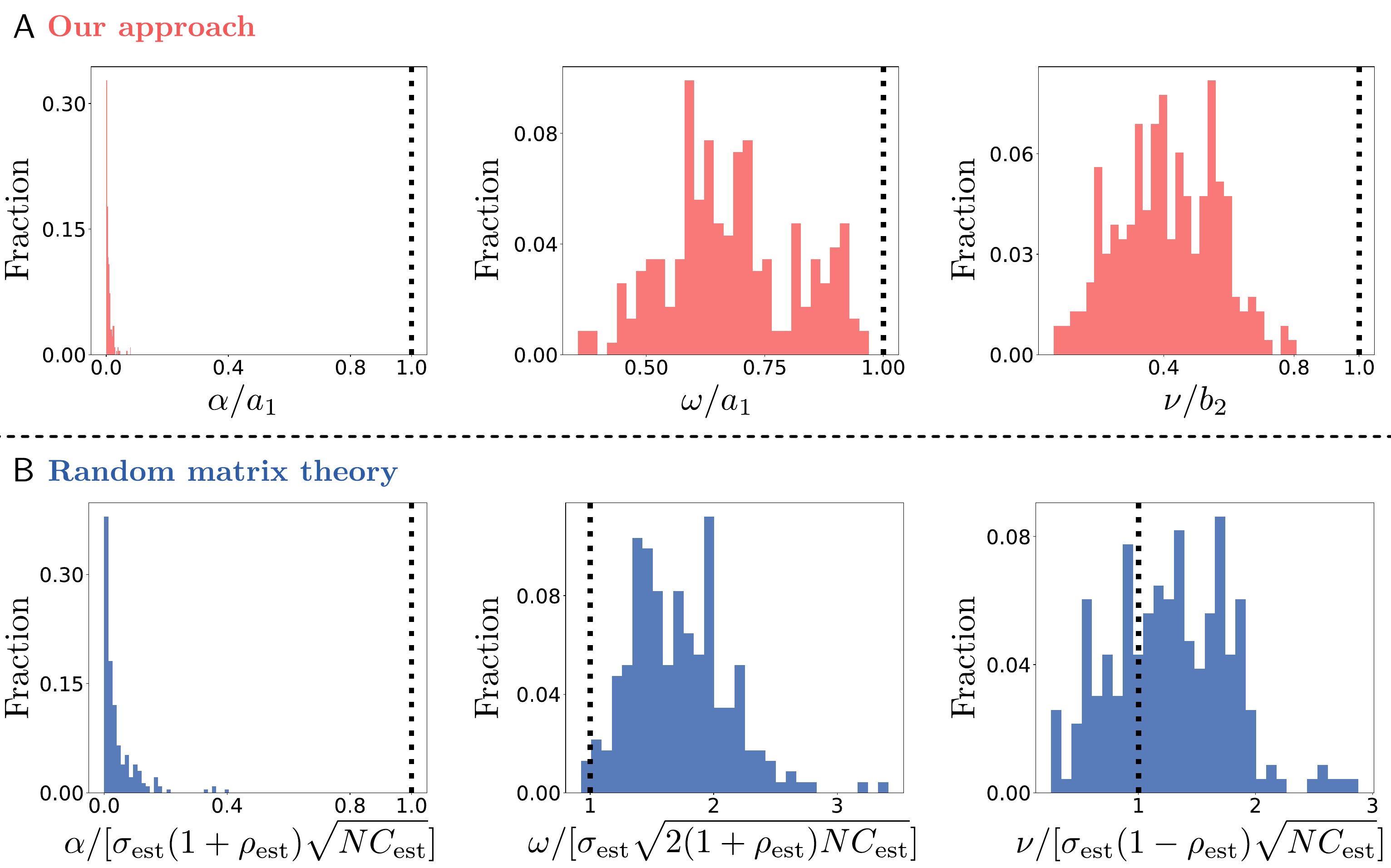}
    \caption{
    Histograms of the ratio between the y-axis and x-axis values in Fig.~4C (A) and Fig.~4E (B).
    The black dotted line marks a ratio of $1$, corresponding to perfect agreement.
    }
    \label{fig:histograms}
\end{figure}

\subsection{Empirical verification of the alternative bounds based on our approach}
We also numerically verified the alternative bounds derived from our approach.
In Fig.~\ref{fig:our_alternative}A and B, we test Eqs.~(12) and (13), i.e., the alternative bounds based on $\widetilde{\M}$.
The overall behavior is similar to that of Eqs.~(5) and (6) in Fig.~4C and Fig.~\ref{fig:histograms}A.
Specifically, all points in Fig.~\ref{fig:our_alternative}A lie below the red solid line, and the $y/x$ ratio is always smaller than $1$.
Thus, Eqs.~(12) and (13) hold for all ecosystems.
The Pearson correlation coefficients were $0.0584$, $0.993$, and $0.716$ for $\alpha$, $\widetilde{\omega}$, and $\nu$ in Fig.~\ref{fig:our_alternative}A, respectively.
Under the null hypothesis that the variables are uncorrelated and normally distributed, the p-values for all three plots in Fig.~\ref{fig:our_alternative}A are smaller than $10^{-3}$, except for $\alpha$ vs. $\widetilde{a}_1$ (p-value $0.376$).
Therefore, Eqs.~(12) and (13) capture the trends in $\widetilde{\omega}$ and $\nu$ for real ecosystems.

In Fig.~\ref{fig:our_alternative}C and D, we test Eqs.~(14) and (15), i.e., the alternative bounds based on $\widehat{\M}$.
The behavior is again similar to that of Eqs.~(5) and (6): Fig.~\ref{fig:our_alternative}C and D show that Eqs.~(14) and (15) hold for real ecosystems.
The Pearson correlation coefficients were $0.225$, $0.984$, and $0.987$ for $\alpha$, $\widehat{\omega}$, and $\nu$ in Fig.~\ref{fig:our_alternative}C, respectively.
The p-values for all three plots in Fig.~\ref{fig:our_alternative}C are smaller than $10^{-3}$.
Therefore, Eqs.~(14) and (15) capture the trends in $\alpha$, $\widehat{\omega}$, and $\nu$ for real ecosystems.

\begin{figure}[htbp]
    \centering
    \includegraphics[width=\linewidth]{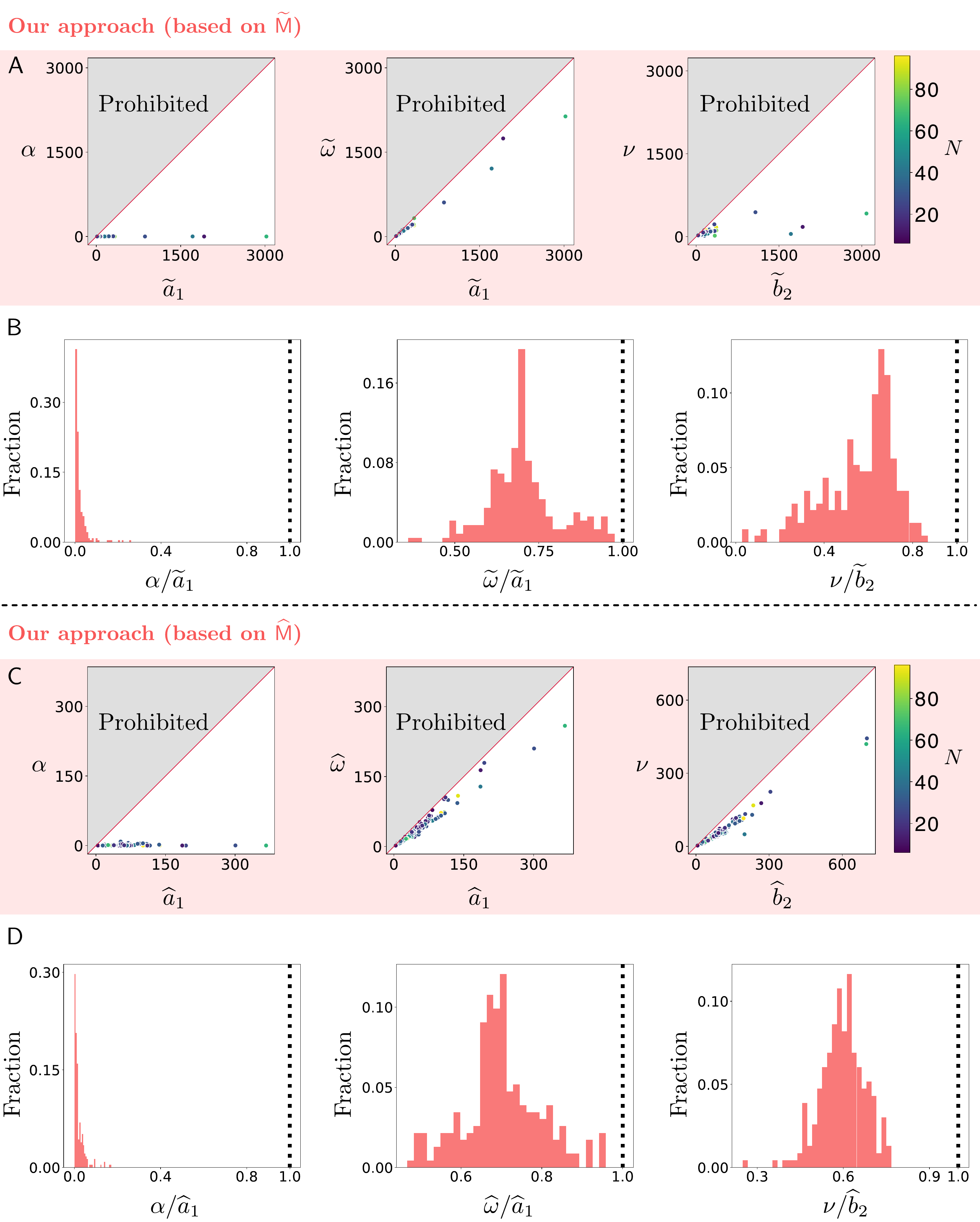}
    \caption{
    (A) Empirical verification of the alternative bounds based on $\widetilde{\M}$ [Eqs.~(12) and (13)].
    The red solid line shows $y=x$.
    No data points lie in the region $y>x$, supporting the validity of Eqs.~(12) and (13).
    (B) Histograms of the $y/x$ ratio for the data in Fig.~\ref{fig:our_alternative}A.
    The black dotted line marks a ratio of $1$, corresponding to perfect agreement.
    (C) Empirical verification of the alternative bounds based on $\widehat{\M}$ [Eqs.~(14) and (15)].
    No data points lie in the region $y>x$, supporting the validity of Eqs.~(14) and (15).
    (D) Histograms of the $y/x$ ratio for the data in Fig.~\ref{fig:our_alternative}C.
    The black dotted line marks a ratio of $1$.
    }
    \label{fig:our_alternative}
\end{figure}

\clearpage

\subsection{Empirical verification of the more general result based on RMT}
In Fig.~\ref{fig:RMT_with_mean}, we empirically tested Eqs.~(\ref{eq:alpha_RMT_general})--(\ref{eq:nu_RMT_general}), which generalize Eqs.~(7)--(9).
The estimators of $\mu^\mathrm{all}$, $\sigma^\mathrm{all}$, and $\rho^\mathrm{all}$ used in Eqs.~(\ref{eq:alpha_RMT_general})--(\ref{eq:nu_RMT_general}) are
\begin{gather}
    \mu^\mathrm{all}_\mathrm{est}:=\frac{1}{N(N-1)}\sum_{i\ne j}M_{ij},\\
    (\sigma^\mathrm{all}_\mathrm{est})^2:=\frac{1}{N(N-1)}\sum_{i\ne j}(M_{ij})^2-\left(\mu^\mathrm{all}_\mathrm{est}\right)^2,\\
    \rho^\mathrm{all}_\mathrm{est}:=\frac{1}{\left(\sigma^\mathrm{all}_\mathrm{est}\right)^2}\left[\frac{1}{N(N-1)/2}\sum_{i>j}M_{ij}M_{ji}-\left(\mu^\mathrm{all}_\mathrm{est}\right)^2\right].
\end{gather}
The behavior of Eqs.~(\ref{eq:alpha_RMT_general})--(\ref{eq:nu_RMT_general}) in Fig.~\ref{fig:RMT_with_mean} is similar to that of Eqs.~(7)--(9) in Fig.~4E and Fig.~\ref{fig:histograms}B.
In other words, Eqs.~(\ref{eq:alpha_RMT_general})--(\ref{eq:nu_RMT_general}) do not match the data well, even for ecosystems with large $N$.
In particular, they tend to overestimate $\alpha$ and underestimate $\omega$, suggesting that RMT does not reproduce the typical magnitudes of these quantities.
The Pearson correlation coefficients were $0.383$, $0.944$, and $0.819$ for $\alpha$, $\omega$, and $\nu$ in Fig.~\ref{fig:RMT_with_mean}A, respectively.
Under the null hypothesis that the variables are uncorrelated and normally distributed, the p-values for all three panels in Fig.~\ref{fig:RMT_with_mean}A are smaller than $10^{-3}$.
These positive and statistically significant correlations indicate that RMT captures the overall trends in $\alpha$, $\omega$, and $\nu$ across ecosystems, even though it does not predict their values accurately.

\begin{figure}[htbp]
    \centering
    \includegraphics[width=\linewidth]{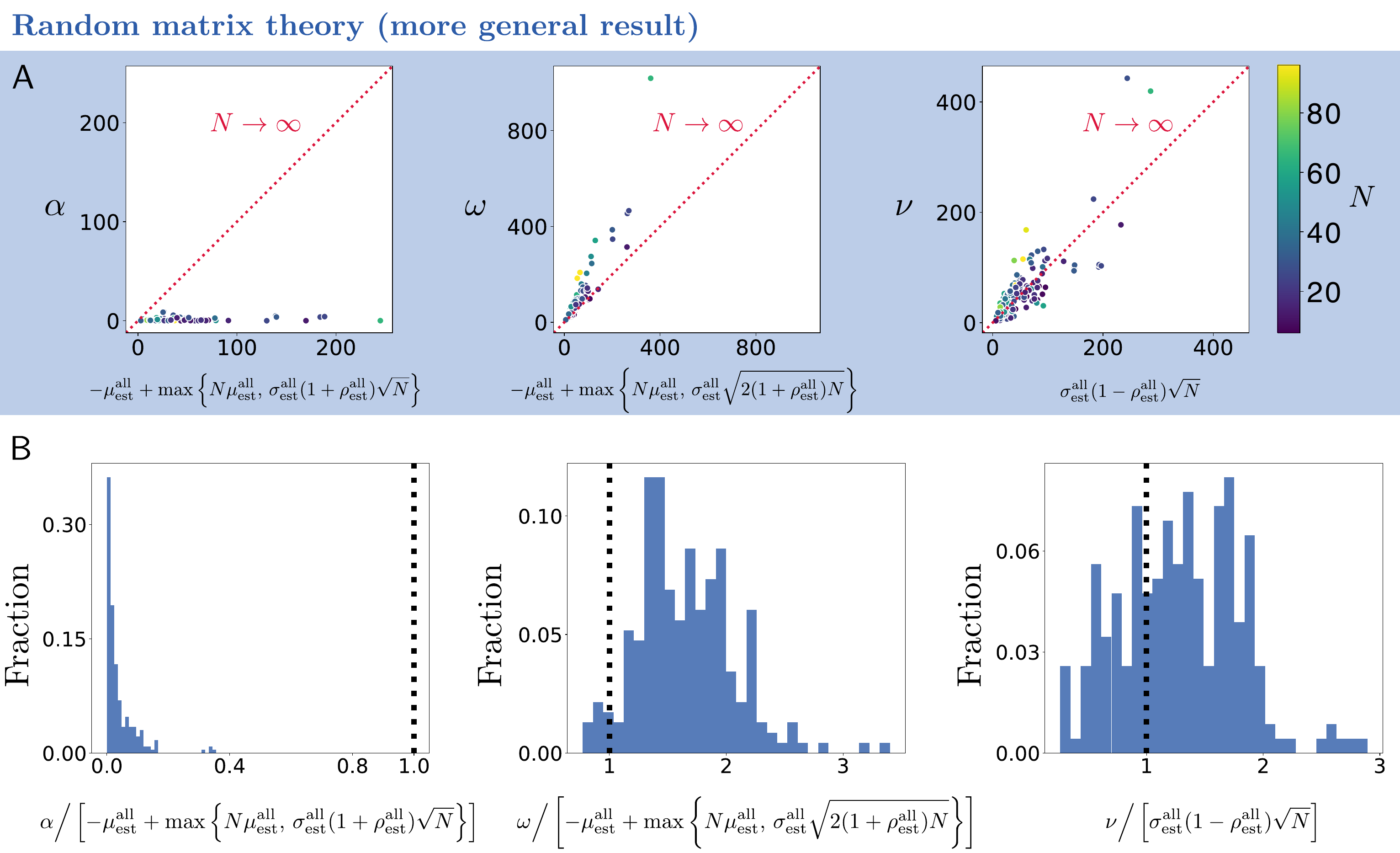}
    \caption{
    (A) Empirical verification of the more general RMT results, Eqs.~(\ref{eq:alpha_RMT_general})--(\ref{eq:nu_RMT_general}).
    The red dotted line shows $y=x$, corresponding to the RMT prediction in the limit $N\to\infty$.
    (B) Histograms of the $y/x$ ratio for the data in Fig.~\ref{fig:RMT_with_mean}A.
    The black dotted line marks a ratio of $1$, corresponding to perfect agreement.
    }
    \label{fig:RMT_with_mean}
\end{figure}

\section{Details of the numerical calculations in the main text}
We describe the details of the numerical calculations in the main text.

\subsection{Details of Fig.1B}
In Fig.1B in the main text, we numerically calculated $\|\delta\bm{x}(t)\|$ from $\frac{d}{dt}\delta\bm{x}(t)=\M\delta\bm{x}(t)$ with
\begin{align}
    \M=
    \begin{pmatrix}
   -1 & 15 \\
   0 & -2
\end{pmatrix},\quad
    \delta\bm{x}(0)=\frac{1}{20}
    \begin{pmatrix}
   1\\
   \sqrt{3}
\end{pmatrix}.
\end{align}
For this $\M$, the spectral abscissa and the numerical abscissa are given by $\alpha=-1$ and $\omega=\frac{\sqrt{226}-3}{2}$, respectively.
In particular, $\alpha<0<\omega$ holds, as mentioned in the main text.

\subsection{Details of Fig.2}
In Fig.2, we numerically calculated the eigenvalues, numerical range, $S_1$, and $S_2$ for a matrix $\M$ given by
\begin{align}
    \M=
    \begin{pmatrix}
   -4.5 & 1    & 0    & 0    & 0    & 0.1   & 0 \\
   -0.5 & -4.3 & 0    & 0    & 0    & 0     & 0 \\
   0    & 0    & -2.5 & 1.1  & 0    & 0     & 0 \\
   0    & 0    & -0.3 & -2.4 & 0    & 0     & 0 \\
   0    & 0    & 0    & 0    & -1.5 & 1     & 0 \\
   0.01 & 0    & 0.01 & 0    & -0.2 & -1.6  & 0.02 \\
   0    & 0    & 0    & 0    & 0    & -0.05 & -2.5 
\end{pmatrix}.
\end{align}
To calculate the numerical range, we used the algorithm in Ref.~\cite{johnson1978numerical}.
For this $\M$, the corresponding values are $\alpha\approx-1.55$, $\omega\approx-1.15$, $\nu\approx0.700$, $a_1\approx0.475$, $b_1=1.425$, $a_2\approx0.972$, $b_2=0.795$, $c^-=-4.5$, and $c^+=-1.5$.

\subsection{Details of Fig.3}
In Fig.3 of the main text, we generated random matrices $\M$ as follows. 
We considered integer values of $N$ from $6$ to $99$ and generated 25 matrices $\M\in\mathbb{R}^{N\times N}$ for each $N$. 
Each diagonal element $(\M)_{ii}$ was independently sampled from the uniform distribution on $[-10,-5]$. 
For each pair $i>j$, with probability $C_N:=20/[N(N-1)]$, the two off-diagonal elements $(\M)_{ij}$ and $(\M)_{ji}$ were independently sampled from the uniform distribution on $[-3,3]$; otherwise, both elements were set to zero. 
With this choice of $C_N$, the expected number of nonzero off-diagonal elements is $N(N-1)C_N=20$, which is independent of $N$.
This choice ensures that the typical magnitudes of $a_1+c^+$ and $b_2$ in Fig.3 do not systematically increase with $N$, because they are mainly controlled by the number and magnitudes of the nonzero off-diagonal elements.

\section{EcoBase data and their references}
We show all models in EcoBase used in this study and their references in Table~\ref{tab:EcoBase}.
Note that a single model listed in the table can include multiple ecosystem data. Therefore, the number of models listed is smaller than the total number of data, 232.

\begin{small}

\tablecaption{EcoBase data used in this study and their references}
\label{tab:EcoBase}

\tablefirsthead{
  \toprule
  \textbf{Model Name} & \textbf{Reference} \\
  \midrule
}
\tablehead{
  \toprule
  \textbf{Model Name} & \textbf{Reference} \\
  \midrule
}
\tabletail{\midrule}
\tablelasttail{\bottomrule}

\begin{supertabular}{p{0.42\linewidth}p{0.50\linewidth}}
Aegean Sea & Keramidas et al. 2022\cite{keramidas2022modelling} \\
Alaska, Prince William Sound & Dalsgaard et al. 1997\cite{dalsgaard1997preliminary} \\
Albatross Bay & Okey 2006\cite{okey2006albatross} \\
Aleutian Islands & Gu{\'e}nette et al. 2006\cite{Guenette2006} \\
Alto Golfo de California & Morales-Zárate et al. 2004\cite{MORALESZARATE2004331} \\
Antarctic & Hoover 2009\cite{hoover2009ecosystem} \\
Apalachicola Bay & Allen 2022\cite{allen2022synergistic} \\
Apalachicola Bay & Allen \& Lewis 2024\cite{ALLEN2024e00372} \\
Arachania & Lercari et al. 2010\cite{LERCARI20102751} \\
Australia North West Shelf & Bulman 2006\cite{bulman2006trophic} \\
Azores & Morato et al. 2016\cite{Morato2016} \\
Azores archipelago & Gu{\'e}nette \& Morato 2001\cite{Guenette2001} \\
Baie de Seine & Halouani et al. 2020\cite{HALOUANI2020103434} \\
Baja California & Cisneros-Montemayor et al. 2012\cite{CISNEROSMONTEMAYOR20128} \\
Bamboung & Coll{\'e}ter et al. 2012\cite{COLLETER20121} \\
Barents Sea & Blanchard et al. 2002\cite{blanchard2002exploring} \\
Barnegat Bay & Vasslides et al. 2017\cite{vasslides2017modeling} \\
Barra Del Chuy & Lercari et al. 2010\cite{LERCARI20102751} \\
Bay of Biscay & Ainsworth et al. 2001\cite{ainsworth2001bay} \\
Bay of Biscay & Moullec et al. 2017\cite{MOULLEC2017104} \\
Bay of Biscay & Lassalle et al. 2012\cite{lassalle2012ecosystem} \\
Black Sea & Akoglu et al. 2014\cite{AKOGLU2014113} \\
Black Sea & Gucu 2002\cite{GUCU2002439} \\
BlackSea & {\"O}rek 2000\cite{orek_2000} \\
Bolinao Coral Reef & Ali{\~n}o et al. 1993\cite{alino1990initial} \\
Bonifacio Straits Natural Reserve & Vanalderweireldt et al. 2024\cite{VANALDERWEIRELDT2024107263} \\
British Columbia coast & Preikshot 2007\cite{Preikshot_2007} \\
Burrard Inlet 1750-1980 CE & Efford 2024\cite{Efford_2024} \\
Calvi Bay & Pinnegar \& Polunin 2004\cite{PINNEGAR2004249} \\
Campeche & Vega-Cendejas et al. 1993\cite{vega1993trophic} \\
Canada, Grand Banks of Newfoundland & Bundy 2002\cite{Bundy2002} \\
Cap de Creus MPA - whole & Corrales et al. 2020\cite{CORRALES2020105232} \\
Cape Verde & Stobberup et al. 2004\cite{Stobberup2004} \\
Caribbean & Melgo et al. 2009\cite{Melgo2009} \\
Carretas-Pereyra coastal lagoon system & L{\'o}pez-Vila et al. 2021\cite{lopez2021trophic} \\
Celestun & Ch{\'a}vez et al. 1993\cite{Chavez1993} \\
Celestun mangrove & Vega-Cendejas \& Arregu{\'{\i}}n-Sánchez 2001\cite{VEGACENDEJAS2001119} \\
Celtic Sea & Hernvann et al. 2020\cite{Hernvann2020} \\
Celtic Sea & Moullec et al. 2017\cite{MOULLEC2017104} \\
Celtic Sea-Biscay & Bentorcha et al. 2017\cite{BENTORCHA2017} \\
Central Atlantic & Vasconcellos \& Watson 2004\cite{Vasconcellos2004} \\
Central Baltic Sea & Tomczak et al. 2012\cite{TOMCZAK2012123} \\
Central Chile & Neira et al. 2004\cite{NEIRA2004233} \\
Central Gulf of California & Arregu{\'{\i}}n-Sánchez et al. 2002\cite{ARREGUINSANCHEZ2002167} \\
Cerbère-Banyuls MPA & Corrales et al. 2020\cite{CORRALES2020105232} \\
Chantuto-Panzacola & L{\'o}pez-Vila et al. 2019\cite{lopez2019young} \\
Chesapeake & Christensen et al. 2009\cite{christensen2009fisheries} \\
Cockburn Sound & Lozano-Montes et al. 2024\cite{lozano-montes2024cockburn} \\
Contemporary Alosine & Dias et al. 2019\cite{Dias2019} \\
Cyprus insular shelf trophic model & Michailidis et al. 2019\cite{MICHAILIDIS2019152} \\
Danajon Bank & Bacalso \& Wolff 2014\cite{BACALSO2014103} \\
Deep Western Mediterranean sea & Tecchio et al. 2013\cite{TECCHIO20131} \\
Denmark, Faroe Islands & Zeller \& Reinert 2004\cite{ZELLER2004403} \\
East Bass Strait & Bulman et al. 2006\cite{bulman2006trophicdynamics} \\
Eastern Bering Sea & Trites et al. 1999\cite{trites1999ecosystem} \\
Eastern Corsican Coast & Vanalderweireldt et al. 2022\cite{VANALDERWEIRELDT2022103798} \\
Eastern Scotian Shelf & Bundy 2004\cite{bundy2004mass} \\
Eritrea & Tsehaye \& Nagelkerke 2008\cite{TSEHAYE2008319} \\
Falkland Islands & Cheung \& Pitcher 2005\cite{Cheung2005} \\
Falkland Islands & Büring et al. 2024\cite{Buring2024} \\
Florida Bay & Smith et al. 2021\cite{smith2021ecosystem} \\
Florida Bay & Smith et al. 2023\cite{Smith2023} \\
Galapagos & Ruiz et al. 2016\cite{RUIZ20161} \\
Galapagos, Floreana rocky reef & Okey et al. 2004\cite{OKEY2004383} \\
Garonne & Palomares et al. 1993\cite{Palomares1993} \\
Georges Bank & Link et al. 2006\cite{link2006emax} \\
Germany, Schlei Fjord & Christensen \& Pauly 1992\cite{CHRISTENSEN1992169} \\
Gironde estuary & Lobry 2004\cite{lobry2004} \\
Golfo Dulce & Wolff et al. 1996\cite{wolff1996pilot} \\
Grand Banks of Newfoundland & Heymans 2003\cite{Heymans2003} \\
Grand Banks of Newfoundland & Heymans \& Pitcher 2002\cite{Heymans2002} \\
Grand Banks region & Wudrick 2023\cite{wudrick2023comparison} \\
Greenland, West Coast & Pedersen \& Zeller 2001\cite{Pedersen2001} \\
Guinea & Gu{\'e}nette \& Diallo 2004\cite{Guenette2004} \\
Guinea & Gascuel et al. 2009\cite{Gascuel2009} \\
Gulf of California & Lercari \& Arregu{\'{\i}}n-Sánchez 2009\cite{Lercari2009} \\
Gulf of Carpentaria & Okey et al. 2007\cite{Okey_2007_GulfCarpentaria} \\
Gulf of Corinth & Raptis et al. 2025\cite{RAPTIS2025104466} \\
Gulf of Gabes & Hattab et al. 2013\cite{HATTAB2013159} \\
Gulf of Maine & Heymans 2001\cite{Heymans2001} \\
Gulf of Maine & Link et al. 2006\cite{link2006emax} \\
Gulf of Mexico & Browder 1993\cite{Browder1993} \\
Gulf of Mexico & Walters et al. 2008\cite{walters2008ecosim} \\
Gulf of Nicoya & Wolff et al. 1998\cite{wolff1998trophic} \\
Gulf of Salamanca & Duarte \& Garc{\'{\i}}a 2004\cite{DUARTE2004323} \\
Gulf of Thailande & Christensen 1998\cite{Christensen1998} \\
Hudson Bay & Hoover 2010\cite{Hoover2010} \\
Huizache-Caimanero & Zetina-Rejón et al. 2003\cite{ZETINAREJON2003803} \\
Humboldt Current & Tam et al. 2008\cite{TAM2008352} \\
Iceland & Buchary 2001\cite{Buchary2001} \\
Icelandic shelf & Mendy 1998\cite{mendy1998trophic} \\
Independence Bay & Taylor et al. 2008\cite{TAYLOR2008336} \\
Irish Sea & Lees \& Mackinson 2007\cite{Lees2007IrishSea} \\
Jalisco and Colima Coast & Galv{\'a}n 2005\cite{Galvan2005} \\
Jurien Bay & Lozano-Montes et al. 2011\cite{Lozano2011} \\
Kaloko Honokohau & Wabnitz et al. 2010\cite{wabnitz2010ecosystem} \\
Kuosheng Bay & Lin et al. 2004\cite{lin2004trophic} \\
Lagoon Chiku - Taiwan & Lin et al. 1999\cite{LIN1999575} \\
Lesser Antilles & Mohammed et al. 2008\cite{Mohammed2008} \\
Liberia & Kay 2011\cite{Kay2011Liberia} \\
Looe Key National Marine Sanctuary & Venier 1997\cite{Venier1997} \\
Low Barents sea & Blanchard et al. 2002\cite{blanchard2002exploring} \\
Madeira archipelago & Romero et al. 2024\cite{ROMERO2024106382} \\
Malangen Fjord & Vilas et al. 2020\cite{VILAS2020103268} \\
Marmara Sea 1990s & Saygu et al. 2023\cite{saygu2023} \\
Marmara Sea 2000s & Saygu et al. 2023\cite{saygu2023} \\
Marmara Sea 2010s & Saygu et al. 2023\cite{saygu2023} \\
Mauritania & Sidi \& Guénette 2004\cite{Sidi2004} \\
Mauritanie & Gu{\'e}nette et al. 2014\cite{guenette2014assessing} \\
Medes Island MPA & Corrales et al. 2020\cite{CORRALES2020105232} \\
Mid-Atlantic Bight & Link et al. 2006\cite{link2006emax} \\
Miramare & Libralato et al. 2006\cite{Libralato2006Miramare} \\
Mississippi River Delta & de Mutsert et al. 2017\cite{DEMUTSERT201714} \\
Moreton Bay Ecosystem & Fondo et al. 2015\cite{fondo2015banning} \\
Morocco & Stanford et al. 2001\cite{Stanford2001} \\
Mount St Michel Bay & Arbach Leloup et al. 2008\cite{ARBACHLELOUP2008111} \\
Narragansett Bay food web & Innes-Gold et al. 2020\cite{innes2020modeling} \\
New Foundland & Ainsworth \& Sumaila 2005\cite{Ainsworth2005} \\
Newfoundland and Labrador shelf region & Wudrick 2023\cite{wudrick2023comparison} \\
Ningaloo & Jones et al. 2011\cite{Jones2011Ningaloo} \\
North Aegean & Tsagarakis et al. 2010\cite{TSAGARAKIS2010233} \\
North Aegean 90s & Tsagarakis et al. 2022\cite{Tsagarakis2022} \\
North Atlantic & Vasconcellos \& Watson 2004\cite{Vasconcellos2004} \\
North Benguela & Watermeyer et al. 2008\cite{Watermeyer01092008} \\
North Brazil & Wolff et al. 2000\cite{WOLFF2000789} \\
North East Pacific & Preikshot 2007\cite{Preikshot_2007} \\
North Sea & Mackinson \& Daskalov 2007\cite{Mackinson2007} \\
North Sea & Christensen et al. 2002\cite{Christensen2002} \\
North Sea & Christensen 1995\cite{Christensen1995NorthSea} \\
North South of China Sea & Cheung 2007\cite{cheung2007vulnerability} \\
North Water Polynya & Bryndum-Buchholz et al. 2024\cite{Bryndum-Buchholz2024} \\
Northern Benguela & Heymans \& Sumaila 2007\cite{heymans2007updated} \\
Northern British Columbia & Ainsworth et al. 2002\cite{Ainsworth2002NorthernBC} \\
Northern Californian Current & Field 2004\cite{Field2004} \\
Northern Californian Current & Walters et al. 2010\cite{Walters2010} \\
Northern Gulf of Mexico & de Mutsert et al. 2025\cite{deMutsert2025} \\
Northern Gulf of Mexico & Sagarese et al. 2017\cite{SAGARESE201775} \\
Northern Gulf of St Lawrence & Savenkoff et al. 2004\cite{Savenkoff2004} \\
Northern Gulf St Lawrence & Morissette et al. 2003\cite{Morissette2003} \\
Northern Humboldt Current & Tam et al. 2008\cite{TAM2008352} \\
Northwest Africa & Morissette et al. 2009\cite{Morissette2009} \\
Orbetello Lagoon & Brando et al. 2004\cite{BRANDO2004213} \\
Pagasitikos Gulf & Dimarchopoulou et al. 2019\cite{Dimarchopoulou2019} \\
Paraná River Floodplain & Angelini \& Agostinho 2005\cite{ANGELINI2005109} \\
Peru & Jarre-Teichmann \& Pauly 1993\cite{JarreTeichmann1993} \\
Peru & Guénette et al. 2008\cite{GUENETTE2008326} \\
Porsangerfjord Area 3 & Pedersen et al. 2018\cite{pedersen2018effects} \\
Port Cros & Valls et al. 2012\cite{valls2012modeling} \\
Port Phillip Bay & Fulton \& Smith 2002\cite{Fulton2002} \\
Portofino & Prato et al. 2016\cite{Prato2016} \\
Prince William Sound & Okey \& Wright 2004\cite{okey2004toward} \\
Raja Ampat & Pitcher et al. 2007\cite{Pitcher2007} \\
Restored Alosine Biomass & Dias et al. 2019\cite{Dias2019} \\
Ria Formosa & Gamito \& Erzini 2005\cite{GAMITO2005509} \\
Ria-Lake Tapajos & Capitani et al. 2021\cite{capitani2021food} \\
Rocas Atoll & Capitani et al. 2022\cite{capitani2022ocean} \\
Santa Pola Bay & Bayle-Sempere et al. 2013\cite{BAYLESEMPERE2013135} \\
Seagrass and Mangrove Terminos Lagoon & Rivera-Arriaga et al. 2003\cite{RiveraArriaga2003} \\
Sechura Bay & Taylor et al. 2008\cite{taylor2008trophic} \\
Senegambia & Samb \& Mendy 2004\cite{Samb2004} \\
Sierra Leone & Heymans \& Vakily 2004\cite{Heymans2004} \\
Sierra Leone 1978 & Heymans \& Vakily 2004\cite{Heymans2004} \\
Sinaloa sur MEXICO & Salcido Guevara 2006\cite{SalcidoGuevara2006} \\
Sirinhaém River & Lira et al. 2018\cite{LIRA201831} \\
Sonda de Campeche & Manickchand-Heileman et al. 1998\cite{MANICKCHANDHEILEMAN1998885} \\
Sonda\_Campeche\_Act & Zetina-Rej{\'o}n \& Arregu{\'i}n-S{\'a}nchez 2002\cite{zetina-rejon2002flujos} \\
South Benguela & Watermeyer et al. 2008\cite{watermeyer2008changes} \\
South Benguela & Shannon et al. 2003\cite{SHANNON200383} \\
South East Alaska & Guénette et al. 2006 \cite{Guenette2006} \\
South of Benguela & Watermeyer et al. 2008\cite{watermeyer2008changes} \\
South Shetlands & Bredesen 2003\cite{bredesen2003krill} \\
South West Viet nam & Christensen et al. 2003\cite{christensen2003fisheries} \\
South western Gulf of Mexico & Arregu{\'{\i}}n-S{\'a}nchez et al. 1993\cite{Arreguin1993} \\
Southern Brazil & Vasconcellos \& Gasalla 2001\cite{VASCONCELLOS2001279} \\
Southern Gulf of St. Lawrence & Savenkoff et al. 2004\cite{Savenkoff2004southern} \\
Southern New England & Link et al. 2006\cite{link2006emax} \\
Sri Lanka & Haputhantri et al. 2008\cite{HAPUTHANTRI2008304} \\
Strait of Georgia & Preikshot 2007\cite{Preikshot_2007} \\
Strait of Gergia & Martell et al. 2002\cite{Martell2002} \\
Sítios Novos reservoir & Bezerra et al. 2018\cite{bezerra2018food} \\
Sørfjord & Falk-Petersen 2004\cite{FalkPetersen2004} \\
Tamiahua & Abarca-Arenas \& Valero-Pacheco 1993\cite{AbarcaArenas1993} \\
Tampa Bay & Chagaris \& Mahmoudi 2010\cite{chagaris2010assessing} \\
Tampa Bay & Walters et al. 2008\cite{walters2008ecosim} \\
Tasmanian Seamounts Marine Reserve & Bulman et al. 2002\cite{bulman2002trophodynamic} \\
Tasmanian waters & Watson et al. 2013\cite{WATSON2013115} \\
Terminos Lagoon & Manickchand-Heileman et al. 1998\cite{manickchand1998energy} \\
Thau & Palomares et al. 1993\cite{Palomares1993trophic} \\
Thermaikos Gulf & Dimarchopoulou et al. 2022\cite{DIMARCHOPOULOU2022107667} \\
Ungava Bay & Tokpanou et al. 2026\cite{tokpanou2026}\\
USA, Mid Atlantic Bight & Okey 2001\cite{Okey2001} \\
USA, South Atlantic Continental Shelf & Okey \& Pugliese 2001\cite{OkeyPugliese2001} \\
Venezuela shelf & Mendoza 1993\cite{Mendoza1993} \\
Virgin Islands & Opitz 1996\cite{opitz1996trophic} \\
West Baffin Bay, Coastal and Shelf & Pedro et al. 2023\cite{pedro2023structure} \\
West Coast of Peninbsular Malaysia & Christensen et al. 2003\cite{christensen2003fisheries} \\
West coast of Sabah & Garces et al. 2003\cite{garces2003trophic} \\
West Coast Vancouver Island & Martell 2002\cite{martell2002variation} \\
West Florida Shelf & Vilas et al. 2020\cite{vilas2020red} \\
West Florida Shelf Historic Model & Chagaris et al. 2015\cite{Chagaris2015} \\
West Scotland & Morissette \& Pitcher 2005\cite{MorissettePitcher2005} \\
West scotland DeepSea & Howell et al. 2009\cite{howell2009deepfish} \\
Western Antarctic Peninsula & Dahood et al. 2020\cite{DAHOOD2020108766} \\
Western Bering Sea & Aydin et al. 2002\cite{aydin2002comparison} \\
Western Channel & Araújo et al. 2005\cite{Araujo2005} \\
Western Tropical Pacific Ocean & Godinot \& Allain 2003\cite{godinot2003preliminary} \\
Yucatan & Arregu{\'{\i}}n-Sánchez et al. 1993\cite{Arreguin1993Yucatan} \\
\end{supertabular}
\end{small}

\FloatBarrier
\bibliographystyleSI{apsrev4-2}
\bibliographySI{biblio_supply}
\end{SIbibliography}

\end{document}